\documentclass[aps, prd, superscriptaddress, preprintnumbers, floatfix, longbibliography, twocolumn, 10pt]{revtex4-2}

\usepackage{graphicx,amsfonts,amsmath,amssymb,amstext}
\usepackage{float,wrapfig}
\usepackage{subfigure, psfrag}
\usepackage{dsfont,bm}
\usepackage{color}
\usepackage[colorlinks=true,urlcolor=blue,citecolor=blue,linkcolor=blue]{hyperref}
\hypersetup{breaklinks=true}
\usepackage{verbatim}
\usepackage{multirow}

\newcommand{\sign}[1]{\,\mbox{sgn}\left({#1}\right)}

\newcommand{\RE}[1]{\,\mbox{Re}\left\{{#1}\right\}}

\usepackage{array}
\newcolumntype{L}[1]{>{\raggedright\let\newline\\\arraybackslash\hspace{0pt}}m{#1}}
\newcolumntype{C}[1]{>{\centering\let\newline\\\arraybackslash\hspace{0pt}}m{#1}}
\newcolumntype{R}[1]{>{\raggedleft\let\newline\\\arraybackslash\hspace{0pt}}m{#1}}

\definecolor{purple}{rgb}{0.8,0,0.6}
\definecolor{darkgreen}{rgb}{0.00,0.6,0.00}

\usepackage[usenames,dvipsnames]{xcolor}

\extrafloats{100}

\begin{document}

\title{Tunneling spectra of impurity states in unconventional superconductors}
\date{July 10, 2023}

\author{P.~O.~Sukhachov}
\email{pavlo.sukhachov@yale.edu}
\affiliation{Department of Physics, Yale University, New Haven, Connecticut 06520, USA}

\author{Felix von Oppen}
%\email{vonoppen@physik.fu-berlin.de}
\affiliation{\mbox{Dahlem Center for Complex Quantum Systems and Fachbereich Physik, Freie Universit\"{a}t Berlin, 14195 Berlin, Germany}}

\author{L.~I.~Glazman}
%\email{leonid.glazman@yale.edu}
\affiliation{Department of Physics, Yale University, New Haven, Connecticut 06520, USA}

\begin{abstract}
We investigate the role of the Bloch functions and superconducting gap symmetries on the formation and properties of impurity-induced resonances in a two-dimensional superconductor, and elucidate their manifestation in scanning tunneling spectra. We use and extend a recently developed scattering approach, conveniently formulating the results in terms of the phase shifts of electron scattering off the impurity. We find that the discrete subgap states in a nodeless-gap superconductor are insensitive to the potential scattering phase shift (common for the two spin species) if time-reversal symmetry (TRS) is preserved. The independence of potential scattering is exact for $s$-wave superconductors. It remains an accurate approximation over a broad range of subgap energies when the gap function breaks the lattice point symmetry, except for a narrow region below the gap edge. Breaking of TRS makes potential scattering capable of creating spin-degenerate subgap states, which may be further split by  spin-dependent scattering. In nodal-gap superconductors, impurity-induced resonances are broadened by coupling to the quasiparticle continuum. We identify the conditions allowing for the formation of narrow resonances. In addition to finding the energy spectrum, we evaluate the spin-resolved differential conductance for all of the considered symmetries and gap structures.
\end{abstract}

\maketitle

\section{Introduction}
\label{sec:Introduction}

Since its invention in 1982~\cite{Binnig-Weibel:1982-STM,Binnig-Rohrer:review-1987}, scanning tunneling microscopy/spectroscopy (STM/STS) has developed into an important tool for investigating properties of a wide range of materials, including conventional and unconventional superconductors~\cite{Deutscher:review-2004,Fischer-Renner:review-2007}. By applying a voltage bias between tip and superconductor, a tunneling current is measured and translated into a differential conductance. In the weak tunneling regime when the tip is away from the sample, the differential conductance quantifies the density of states (DOS) of the superconductor, provided that the tip's DOS is featureless. While a U-shaped profile of the differential conductance is a signature of a nodeless gap as in conventional $s$-wave superconductors, a V-shaped profile is widely attributed to a nodal superconducting gap as, e.g., in $d$-wave superconductors. The observation of a V-shaped differential conductance was used to claim unconventional pairing in high-$T_c$ superconductors such as, e.g., Bi$_2$Sr$_2$CaCu$_2$O$_{8+\delta}$~\cite{Renner-Fischer:1995} and YBa$_2$Cu$_3$O$_{7-\delta}$~\cite{Maggio-Aprile-Erb:1995}.

Recently, there has been a new surge of interest in unconventional superconductors related to the experimental realization of moir\'{e} heterostructures; see Ref.~\cite{Andrei-MacDonald:2020-rev} for a review. In particular,  superconductivity was reported in twisted bilayer (TBG), trilayer (TTG), and multilayer graphene~\cite{Cao-Jarillo-Herrero:2018-TBG,Yankowitz-Dean:2018,Lu-Efetov:2019,Cao-Jarillo-Herrero:2020,Oh-Yazdani:2021,Kim-Nadj-Perge:2021,Battista-Efetov:2021,Zhang-Nadj-Perge:2021}. STS experiments on TBG and TTG~\cite{Oh-Yazdani:2021,Kim-Nadj-Perge:2021} observed a V-shaped profile of the differential conductance in the weak tunneling regime. Combined with an enhanced low-bias conductance in the strong-tunneling regime~\cite{Oh-Yazdani:2021}, this was interpreted as a signature of nodal superconductivity. The greater tunability of TBG and TTG compared to high-$T_c$ superconductors makes moir\'{e} heterostructures a promising platform for investigating unconventional superconductivity. A complete understanding of the symmetry of the superconducting gap in TBG and TTG, however, remains elusive.

Signatures of the symmetry of the superconducting order parameter are also provided by investigating how the local density of states (LDOS) is modified by defects or impurities; see Refs.~\cite{Balatsky-Zhu:rev-2006,Alloul-Hirschfeld:review-2007} for reviews. Scattering off an impurity may result in the appearance of narrow peaks in the differential conductance as a function of the applied voltage bias, indicative of bound states within a nodeless gap and resonances within a nodal gap. For $s$-wave superconductors, bound states are induced only by scattering potentials that break time-reversal symmetry (TRS) and are known as Yu-Shiba-Rusinov (YSR) states~\cite{Yu:1965,Shiba:1968,Rusinov:1968,Rusinov:1969}. Such a TRS-breaking potential is induced, e.g., by magnetic adatoms, and the corresponding YSR states were probed in Refs.~\cite{Yazdani-Eigler:1997,Ji-Xue:2008-YSR,Ji-Ma:2010-YSR,Franke-Pascual:2011}; see also Ref.~\cite{Heinrich-Franke:2017-review} for a review. Unlike YSR states, resonances in nodal gaps do not necessarily require a TRS-breaking scattering potential~\cite{Hirschfeld-Einzel:1988,Balatsky-Rosengren:1995,Salkola-Scalapino:1996,Salkola-Schrieffer:1997} and can be seen as asymmetric peaks in the conductance~\cite{Yazdani-Eigler:1999,Pan-Davis:2000}. Information on the symmetry of the gap can also be gleaned from the LDOS pattern around the impurity; see, e.g., Refs.~\cite{Salkola-Scalapino:1996,Salkola-Schrieffer:1997} for a theoretical analysis. A four-fold impurity-induced LDOS pattern aligned with the nodes of a $d$-wave superconducting gap was observed via STM in Refs.~\cite{Pan-Davis:2000,Hudson-Davis:2001}. In addition to the symmetry of the gap, the spatial profile of a YSR state may also provide information about the shape of the Fermi surface as was exemplified for an $s$-wave superconductor in Ref.~\cite{Uldemolins-Simon:2022}.

Theoretically, the formation of impurity states in superconductors is typically studied by directly solving the Bogoliubov-de Gennes equations~\cite{Rusinov:1968,Oppen-Pientka:2017}, applying the T-matrix formalism~\cite{Shiba:1968,Hirschfeld-Einzel:1988,Balatsky-Zhu:rev-2006}, or using the numerical renormalization group~\cite{Shiba-Shimizu:1993,Bauer-Hewson:2007}. While being powerful, these approaches can be cumbersome and, for the most part, do not take the structure of the Bloch functions into account. Recently, we proposed an alternative approach to the differential conductance in an STM setting~\cite{Sukhachov-Glazman:2022-STM}. Our method employs the scattering matrix formalism to describe point-contact tunneling into a superconductor, accounting for  arbitrary symmetries of the gap and the Bloch functions as well as arbitrary transmission coefficients between tip and superconductor. Scattering theory accounts for the single-particle and Andreev-reflection contributions to the conductance in a unified and intuitive way. An additional benefit of the scattering framework is that it readily accounts for the scattering potential induced by the tip through the appropriate \emph{phase} of the contact's scattering matrix.

In this work, we apply and extend this framework~\cite{Sukhachov-Glazman:2022-STM} to describe impurity-induced subgap states and resonances in superconductors. Considering two-dimensional (2D) superconductors, we focus on the role of the symmetries of the superconducting order parameter and of the Bloch functions. To this end, we extend the framework of Ref.~\cite{Sukhachov-Glazman:2022-STM} to scattering off an impurity with arbitrary scattering potential, as described in terms of phase shifts, while remaining in the regime of weak tunneling between STM tip and superconductor. For nodeless gaps preserving TRS and lattice point symmetry, the energy of YSR states depends only on the spin-dependent scattering phase. This independence on the potential-scattering phase shift becomes approximate but still accurate in a broad subgap energy range for superconductors breaking the point symmetry. Deviations appear near the gap edges, where the potential-scattering phase shift affects the nucleation of subgap states. In a superconductor with broken TRS, potential scattering can induce spin-degenerate subgap states. The spin degeneracy is lifted when the scattering phase shifts become spin dependent. Resonance states in nodal-gap superconductors exist for any type of scattering but are broadened by the states' ``leakage'' into the quasiparticle continuum. If the nodal gap preserves the lattice point symmetry, the resonances are narrow and located near the Fermi level for strong scattering. In the absence of the point symmetry, the appearance of low-energy resonances requires fine-tuning. Along with the energy spectrum of the resonances, we evaluate the spin-resolved differential conductance for all of the considered symmetries and gaps. While the energies of the resonances are always particle-hole symmetric, their strengths as manifested in the peak heights in the differential conductance develop strong asymmetry for nodal gaps. This asymmetry is explained by the difference between the conductance determined by the full scattering amplitudes and the particle-hole-symmetric density of energy levels. Thus, our results provide complementary information about the symmetry of the gap and the Bloch functions as well as the properties of the impurity. Experimentally, the spin-resolved differential conductance can be mapped via spin-polarized STM/STS~\cite{Schneider-Wiesendanger:2020,Wang-Wiesendanger:2020}; see also Ref.~\cite{Wiesendanger:review-2009} for a review of the technique.

This paper is organized as follows. We present the extended scattering framework in Sec.~\ref{sec:Model}. The bound states for nodeless gaps are analyzed in Sec.~\ref{sec:Bound} for several symmetries of the gap. Section~\ref{sec:Resonances} is devoted to resonance states in nodal-gap superconductors. The results are summarized and discussed in Sec.~\ref{sec:Summary}. Technical details of the derivation of scattering amplitudes and approximate results for low-energy states are given in Appendices~\ref{sec:App-Model} and \ref{sec:app-small-bias}, respectively. The dispersion relation of the bound states in a fully-gapped $d_{x^2-y^2}+id_{xy}$ superconductor and the differential conductance for above-gap energies are presented in Appendices~\ref{sec:App-d+id} and \ref{sec:App-Resonances-above}, respectively. Throughout this paper, we assume zero temperature.

\section{Scattering theory for STS spectra}
\label{sec:Model}

The hybridization of impurity atom and substrate electrons introduces both exchange and potential scattering. Placing the tip above the impurity, the impurity spin is further hybridized with the tip electrons. A Schrieffer-Wolff transformation eliminating empty and doubly-occupied impurity orbitals leads to the effective coupling
\begin{equation}
\label{Model-Hint-def}
H_\mathrm{int} = \sum_{\mathbf{k},\mathbf{k'}}\sum_{\alpha,\beta} \psi^\dagger_{\alpha,\mathbf{k},\sigma} [V_{\alpha\beta} \delta_{\sigma\sigma'} + J_{\alpha\beta}\mathbf{S}\cdot\mathbf{s}_{\sigma\sigma'}] \psi^{\phantom{\dagger}}_{\beta,\mathbf{k}',\sigma'}.
\end{equation}
Here, $\alpha,\beta\in \{L,R\}$ refer to the tip (L) and the substrate (R) electrons, $\sigma=\uparrow,\downarrow$ denotes the spin projections, $\mathbf{s}$ is the electron spin operator, and $\psi_{\alpha,\mathbf{k},\sigma}$ annihilates an electron in the lead $\alpha$, with the momentum $\mathbf{k}$, and the spin $\sigma$. The first term involving $V_{\alpha\beta}$ represents potential coupling, the second term involving $J_{\alpha\beta}$ is the exchange coupling with the impurity spin $\mathbf{S}$.

We treat the impurity spin as classical, $\mathbf{S}=S\mathbf{\hat z}$~\footnote{The classical model of impurity-tip contact is valid for large easy-axis anisotropy. A detailed discussion of magnetized impurity in classical and quantum approximations can be found in Ref.~\cite{VonOppen-Franke:2021}.}. Then, for each spin polarization $\sigma$, the tip-substrate contact can be described by a two-channel scattering matrix, with one channel in the tip and the other in the substrate,
\begin{equation}
\label{Model-S-def}
\hat{s}_{\sigma}(\varepsilon) =\left(
  \begin{array}{cc}
    s_{\sigma}^{\prime}(\varepsilon) & t_{\sigma}^{\prime}(\varepsilon) \\
    t_{\sigma}(\varepsilon) & s_{\sigma}(\varepsilon) \\
  \end{array}
\right).
\end{equation}
In writing Eq.~(\ref{Model-S-def}), we used the locality of the contact between the tip and the 2D material, so that the contact can be considered as a single-mode quantum point contact opening into the 2D material. The impurity-induced formation of YSR states or resonances results from the scattering matrix element $s_{\sigma}(\varepsilon)$ describing  scattering between in- and outgoing particle waves in the 2D system. The transmission matrix elements $t_{\sigma}^{\prime}(\varepsilon)$ and $t_{\sigma}(\varepsilon)$ determine the transfer rates from the tip into the superconductor and vice versa. Finally, the scattering matrix element $s_{\sigma}^{\prime}(\varepsilon)$ at energy $\varepsilon$ describes reflection between incoming and outgoing channels modeling the STM tip.

In the absence of tunneling, scattering within the 2D material has $|s_{\sigma}(\varepsilon)|=1$, but the phase of $s_{\sigma}(\varepsilon)$ is nontrivial due to electron scattering off the impurity. For a point-like impurity, these properties are encoded in the scattering phases $\delta_\sigma$, which are in general dependent on the spin direction $\sigma$ due to the exchange coupling~\cite{Rusinov:1968},
\begin{equation}
\label{Model-impurity-phase}
\tan{\delta_{\sigma}} = -\pi \nu_0 \left(V_{RR}-\sigma J_{RR} S\right).
\end{equation}
Here $\nu_0$ is the normal-state DOS at the Fermi level. It will be useful to introduce the scattering phase shifts $\delta_s=\delta_{\uparrow}-\delta_{\downarrow}$ and $\delta_c=\delta_{\uparrow}+\delta_{\downarrow}$, where $\delta_s$ describes spin-dependent and $\delta_c$ potential scattering by the impurity. These phase shifts are related, respectively, to the spin and charge of the impurity by the Friedel sum rule; $\delta_s\in[-\pi,\pi]$, $\delta_c\in [0,2\pi]$.

Electron tunneling into the tip reduces the scattering amplitude, $|s_{\sigma}(\varepsilon)|^2=1-|t_\sigma(\varepsilon)|^2$, and also introduces corrections $\propto |t_\sigma(\varepsilon)|^2$ into the scattering phases $\delta_\sigma$. We assume that tunneling and image potential induced by the STM tip lead only to weak perturbations. Accounting for nonzero values of $|t_\sigma(\varepsilon)|^2$ is important in the evaluation of the tunneling current, as the latter is $\propto |t_\sigma(\varepsilon)|^2$. Tunneling also contributes to the width of peaks in tunneling spectra. Such contributions are $\propto |t_\sigma(\varepsilon)|^2$. We will dispense, however, with the effect of tunneling on $\delta_\sigma$. The latter effect leads to shifts of the spectral peaks which are of the order of or smaller than the peak widths (for further discussion of this effect, see Sec.~\ref{sec:Bound}). We will also disregard possible effects of the image potential, which can be minimized by a proper choice of tip material~\cite{Yazdani-privatecomm}.

For contacts between a normal-metal tip and a 2D superconductor, it is convenient~\cite{Beenakker:1992,Nazarov-Blanter:book} to extend the scattering matrix (\ref{Model-S-def}) to Nambu space by introducing scattering matrices for particles, $\hat{s}_{p,\sigma}(\varepsilon)=\hat{s}_{\sigma}(\varepsilon)$, and holes, $\hat{s}_{h,-\sigma}^{*}(-\varepsilon)=\hat{s}_{\sigma}(\varepsilon)$. Then, by using the particle-hole symmetry, one can consider only positive energies $\varepsilon>0$. In what follows, we neglect the energy dependence of $\hat{s}_{\sigma}(\varepsilon)$, assuming it to be featureless on the energy scale of the superconducting gap,
\begin{equation}
\label{Model-sp}
s_{p,\sigma} = s_{0}e^{2i\delta_{\sigma}}, \,\, s_{h,-\sigma} = s_{0} e^{-2i\delta_{-\sigma}},\,\, |t_{p,\sigma}|=|t_{h,\sigma}|\equiv|t|.
\end{equation}
As discussed above, at weak tunneling $1-|s_0|^2=|t|^2\ll 1$, the phases $\delta_\sigma\in [0,\pi]$ encode the electron scattering off an isolated impurity in a 2D material. Notice that the spin dependence of the scattering matrix can originate from a magnetic impurity or a magnetized tip. Thus, we do not require TRS of the normal state of the contact~\footnote{For a magnetic tip, we assume, however, that the magnetization is collinear with that of the magnetic impurity}.

For an impurity embedded in a uniform system, $s_{p,\sigma}$ describes scattering in the zero-angular-momentum channel. An incoming wave $\psi^{\rm in}$ is scattered into the outgoing wave
\begin{equation}
\label{Model-psi-out}
\psi^{\rm out}=\left[(\hat{I}-\hat{P})+s_{p,\sigma}\hat{P}\right]\psi^{\rm in}
\end{equation}
with $\hat P$  being the projector onto zero angular momentum. While the zero-angular-momentum component of the incoming wave is scattered by the contact (term proportional to $\hat P$), nonzero angular momenta remain unaffected (term proportional to $\hat I-\hat P$). In a 2D crystal, angular momentum is no longer a good quantum number, and the angular distribution at the position ${\bf r}_0$ of the impurity is governed by the Bloch function $u_{\bf k}({\bf r}_0)$. The wave vectors ${\bf k}$ at a given energy $\varepsilon$ are defined by $\xi({\bf k})=\varepsilon$ with $\xi({\bf k})$ denoting the dispersion relation measured from the Fermi energy. The projection operator then reads
\begin{equation}
\label{Model-P-def}
\hat{P}\psi^{\rm in}_{\mathbf{k}} = u_{\mathbf{k}}({\bf r}_0)\langle u_{\mathbf{k}^\prime}^*({\bf r}_0)\psi^{\rm in}_{\mathbf{k}^\prime}\rangle_\varepsilon
\end{equation}
with a properly normalized $u_{\mathbf{k}}({\bf r}_0)$ and $\langle\dots\rangle_\varepsilon$ denoting averaging over the constant-energy contour.

\begin{figure}[t]
\centering
\subfigure{\includegraphics[width=0.45\textwidth]{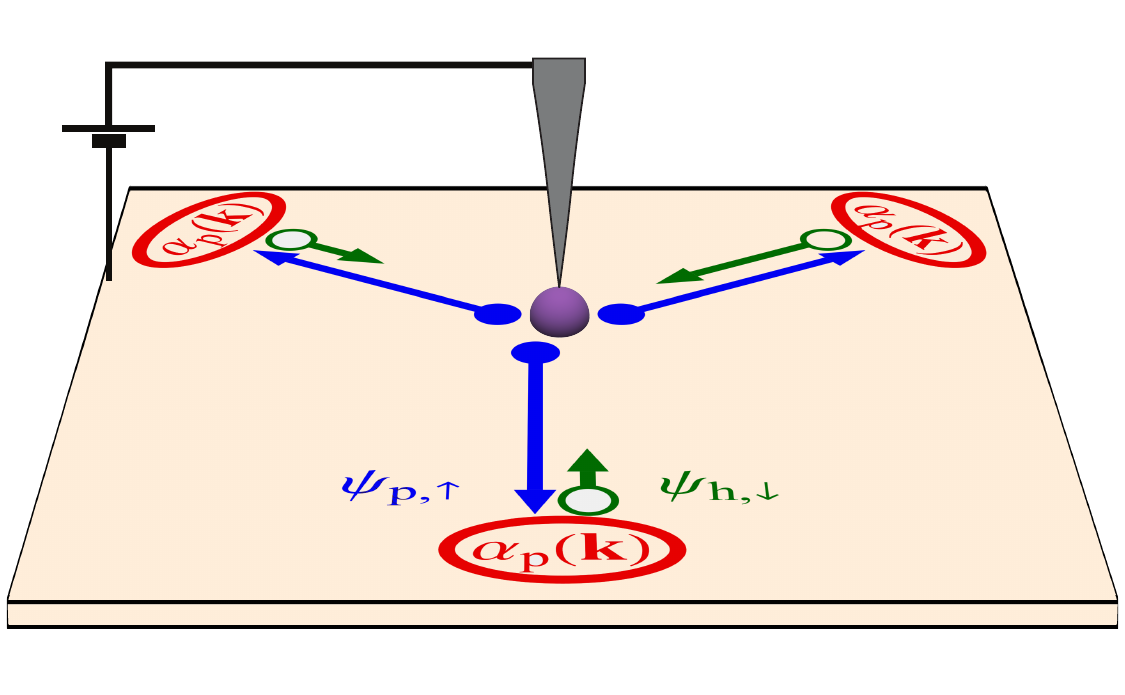}}
\caption{
Schematic setup with an STM tip placed above an impurity in a 2D superconductor. The symmetric blue arrows denote the particle wave emitted by the tip, which carries information about the symmetry of crystalline lattice as encoded in the Bloch function. Asymmetric green arrows represent the Andreev-reflected hole wave that carries information about the superconducting gap symmetry. Bound-state and resonance formation result from repeated cycles of scattering from superconductor and impurity.}
\label{fig:Model-setup}
\end{figure}

To study the transport properties of the contact, we generalize the steps outlined in Ref.~\cite{Sukhachov-Glazman:2022-STM}, where potential scattering off the tip was considered, see Fig.~\ref{fig:Model-setup} for a schematic setup. We relegate details to App.~\ref{sec:App-Model} and focus on the main line of argument here. In the scattering framework, an electron tunneling into a 2D material forms an expanding particle wave with amplitude and directional profile determined by the Bloch function $u_{\mathbf{k}}(\mathbf{r}_0)$, see the blue arrows in Fig.~\ref{fig:Model-setup}. The superconductor retroreflects the particle wave into a counter-propagating hole wave~\cite{Andreev:1964} that carries information about the symmetry of superconducting order parameter, see the green arrows in Fig.~\ref{fig:Model-setup}. Formally, this information enters through the Andreev retroreflection amplitude $\alpha_{p}(\mathbf{k},\varepsilon)$. In the eikonal approximation, which is valid when the coherence length of the superconductor is large compared to the Fermi wavelength, the Andreev retroreflection amplitude $\alpha_{p}(\mathbf{k},\varepsilon)$ depends on the superconducting gap $\Delta(\mathbf{k})$ at the same wave vector $\mathbf{k}$. This allows us to use the result of Refs.~\cite{Beenakker:1992,Nazarov-Blanter:book} at each $\mathbf{k}$, see Eq.~(\ref{Model-alpha-def}) for the explicit expression for $\alpha_{p}(\mathbf{k},\varepsilon)$.

Since the tunneling tip absorbs only a small portion of the returning wave, one needs to sum over many cycles in which the impinging particles undergo Andreev scattering processes and scatter off the contact. This multiple scattering leads to the formation of the YSR bound states or resonances, which are probed by the tip. For a spin-singlet superconductor, the resulting Andreev- and normal-reflection amplitudes of particles incident from the tip are (see Appendix~\ref{sec:App-Model} for details)
\begin{eqnarray}
\label{Model-rph}
r_{ph,\sigma} &=& t_{h,-\sigma}^{\prime} t_{p,\sigma} \frac{a_p}{\mathcal{D}_{\sigma}},\\%\quad \mbox{and} \quad
\label{Model-rp}
r_{p,\sigma} &=& s_{p,\sigma}^{\prime} + t_{p,\sigma}^{\prime} t_{p,\sigma} \frac{a_{ph} +\left(1-s_{h,-\sigma}\right) \left(a_{ph}^2 -a_{p}a_{h}\right)}{\mathcal{D}_{\sigma}},\nonumber\\
\end{eqnarray}
where we use the notation
\begin{eqnarray}
\label{Model-ap-def}
a_{p} &=& \frac{1}{2i} \left\langle  \frac{|u_{\mathbf{k}}(\mathbf{r}_0)|^2 \,\Delta^{*}(\mathbf{k})}{\sqrt{|\Delta(\mathbf{k})|^2-\varepsilon^2}} \right\rangle_\varepsilon,\\
\label{Model-ah-def}
a_{h} &=& \frac{1}{2i}\left\langle \frac{|u_{\mathbf{k}}(\mathbf{r}_0)|^2 \Delta(\mathbf{k})}{\sqrt{|\Delta(\mathbf{k})|^2-\varepsilon^2}} \right\rangle_\varepsilon,
\end{eqnarray}
\begin{equation}
\label{Model-aph-def}
a_{ph} = \frac{1}{2i}\left\langle |u_{\mathbf{k}}(\mathbf{r}_0)|^2 \left[\frac{\varepsilon}{\sqrt{|\Delta(\mathbf{k})|^2-\varepsilon^2}} - i\right] \right\rangle_\varepsilon.
\end{equation}
We also introduced the shorthand
\begin{eqnarray}
\label{Model-D-def}
\mathcal{D}_{\sigma} &=& 1 + \left(2-s_{p,\sigma}-s_{h,-\sigma}\right)a_{ph} \nonumber\\ &+&\left(1-s_{p,\sigma}\right)\left(1-s_{h,-\sigma}\right) \left(a^2_{ph} -a_p a_h\right)
\end{eqnarray}
for the denominator arising from multiple scattering of substrate electrons and holes at the impurity. The analytical continuation to $\varepsilon>|\Delta(\mathbf{k})|$ is defined as $\sqrt{-z}=-i\sqrt{|z|}$ at $z>0$. The averaging $\langle|u_{\mathbf{k}}({\bf r}_0)|^2\dots\rangle_\varepsilon$ over the constant-energy contour in Eqs.~(\ref{Model-ap-def}), (\ref{Model-ah-def}), and (\ref{Model-aph-def}) takes the explicit form
\begin{equation}
\label{Model-average-def}
\left\langle \left|u_{\mathbf{k}}({\bf r}_0)\right|^2\dots\right\rangle_\varepsilon= \frac{\int d^2k\,\delta (\xi({\mathbf{k}})-\varepsilon)\left|u_{\mathbf{k}}({\bf r}_0)\right|^2\dots}{\int d^2k\, \delta (\xi({\mathbf{k}})-\varepsilon)\left|u_{\mathbf{k}}({\bf r}_0)\right|^2}.
\end{equation}
The scattering amplitudes for holes $r_{hp, \sigma}$ and $r_{h, \sigma}$ are obtained  by replacing $a_p\leftrightarrow a_h$, $s_{p,\sigma}\leftrightarrow s_{h,\sigma}$, $s_{p,\sigma}^{\prime}\leftrightarrow s_{h,\sigma}^{\prime}$, $t_{p,\sigma}\leftrightarrow t_{h,\sigma}$, and $t_{p,\sigma}^{\prime}\leftrightarrow t_{h,\sigma}^{\prime}$ in Eqs.~(\ref{Model-rph}) and (\ref{Model-rp}).

Similar to Ref.~\cite{BTK:1982}, the spin-resolved differential conductance $G_{\sigma}(V,\mathbf{r}_0) = {dI_{\sigma}(V,\mathbf{r}_0)}/{dV}$ as a function of the voltage bias $V$, the position of the tip $\mathbf{r}_0$, and the spin projection $\sigma$ reads
\begin{widetext}
\begin{equation}
\label{Resonances-gen-G-def}
G_{\sigma}(V,\mathbf{r}_0)= G_Q \left\{1+ \left[\left|r_{ph,\sigma}(eV,\mathbf{r}_0)\right|^2-\left|r_{p,\sigma}(eV,\mathbf{r}_0)\right|^2\right]\theta(V) +\left[\left|r_{hp,\sigma}(-eV,\mathbf{r}_0)\right|^2-\left|r_{h,\sigma}(-eV,\mathbf{r}_0)\right|^2\right]\theta(-V) \right\}.
\end{equation}
\end{widetext}
This expression reflects that two electrons are transferred from tip to substrate in an Andreev reflection, while no electrons are transferred for normal reflections. The scattering amplitudes in Eqs.~(\ref{Model-rph}) and (\ref{Model-rp}) are used for positive voltage bias $V>0$; see the discussion after Eq.~(\ref{Model-average-def}) for the definition of $r_{hp,\sigma}(-eV,\mathbf{r}_0)$ and $r_{h,\sigma}(-eV,\mathbf{r}_0)$. The conductance quantum is $G_Q=e^2/(2\pi \hbar)$. In the following, we omit the explicit arguments $V$ and $\mathbf{r}_0$ if this does not lead to confusion.

The conductance for arbitrary symmetries of the superconducting gap and the Bloch functions is straightforwardly obtained by substituting Eqs.~(\ref{Model-rph})--(\ref{Model-average-def}) into Eq.~(\ref{Resonances-gen-G-def}). As the final expressions are cumbersome, we focus on representative analytical results in Secs.~\ref{sec:Bound} and \ref{sec:Resonances}. To facilitate the calculations, we find it convenient to rewrite the square of the absolute value of the scattering amplitudes given in Eqs.~(\ref{Model-rph}) and (\ref{Model-rp}) as
\begin{equation}
\label{Resonances-gen-rp2}
\left|r_{p,\sigma}\right|^2 = \frac{\left|-s_{p,\sigma}^{*}\mathcal{D}_{\sigma}+|t|^2\mathcal{N}_{p,\sigma}\right|^2}{|\mathcal{D}_{\sigma}|^2}, \quad \left|r_{ph,\sigma}\right|^2 = |t|^4 \frac{|a_p|^2}{|\mathcal{D}_{\sigma}|^2},
\end{equation}
where we used $s^{\prime}_{p,\sigma} = -s_{p,\sigma}^{*} t_{p,\sigma}/(t_{p,\sigma}^{\prime})^{*}$ and $|t_{p,\sigma}|=|t_{p,\sigma}^{\prime}|=|t|$, which follow from the unitarity of the scattering matrix (\ref{Model-S-def}). Similar expressions can be written for the scattering amplitudes of holes $\left|r_{h,\sigma}\right|^2$ and $\left|r_{hp,\sigma}\right|^2$, see also the replacement rules after Eq.~(\ref{Model-average-def}). The term in the numerator is
\begin{equation}
\label{Resonances-gen-pha-s-Np}
\mathcal{N}_{p,\sigma} = a_{ph} + \left(1 -s_{h,-\sigma}\right) \left(a_{ph}^2 -a_p a_h\right)
\end{equation}
and $\mathcal{N}_{h,\sigma}$ is obtained by replacing $s_{h,-\sigma} \to s_{p,-\sigma}$.

In what follows, we apply this scattering framework to investigate the formation and properties of bound states as well as resonances in the setup, in which the STM tip is placed over the impurity, see also Fig.~\ref{fig:Model-setup}, and discuss the corresponding differential conductance. According to the general tenets of scattering theory~\cite{Baz-Zeldovich-Perelomov:book}, the energies of bound states and resonances correspond to the poles of the scattering amplitudes $r_{ph,\sigma}$ and $r_{p,\sigma}$, which are given in Eqs.~(\ref{Model-rph}) and (\ref{Model-rp}), respectively. Moreover, the differential conductance (\ref{Resonances-gen-G-def}) also contains information  about the eigenfunctions of the impurity-induced states, which, as we explain in Sec.~\ref{sec:Resonances}, can have characteristic symmetry properties.

We consider the effects of impurity scattering for an arbitrary scattering potential in two characteristic regimes: (i) subgap energies $\varepsilon<\Delta_{\rm min} \equiv \mbox{min}{\left\{|\Delta(\mathbf{k})|\right\}}$ for nodeless gaps and (ii) intermediate energies $0<\varepsilon<\Delta_{\rm max} \equiv \mbox{max}{\left\{|\Delta(\mathbf{k})|\right\}}$ for nodal gaps. (The case of above-gap energies $\varepsilon>\Delta_{\rm max}$ is briefly addressed in Appendix~\ref{sec:App-Resonances-above}.) In each of these regimes, we pay special attention to the role of time-reversal and lattice point symmetries.

\section{Bound states within a nodeless gap}
\label{sec:Bound}

We start by considering  nodeless gaps and subgap energies $\varepsilon<\Delta_{\rm min}$. To determine the position of the bound states, we look for zeros of the denominator of the scattering amplitudes, $\mathcal{D}_{\sigma}(\varepsilon)=0$. For subgap energies, it is important to keep terms of first order in $|t|^2$ in the denominator, which determine the width of the peaks in the differential conductance and are thus crucial for describing the resonant tunneling into a superconductor. The zero- and first-order terms in $\mathcal{D}_{\sigma}(\varepsilon) \approx e^{i\sigma \delta_s} \left[\mathcal{D}_{\sigma}^{(0)}(\varepsilon) + i|t|^2 \mathcal{D}_{\sigma}^{(1)}(\varepsilon)\right]$ read
\begin{widetext}
\begin{eqnarray}
\label{Resonances-sub-D-0}
\mathcal{D}_{\sigma}^{(0)}(\varepsilon)
&=& \cos{\delta_s}+\cos{\delta_c}- \sigma \sin{\delta_s} \left[I_{+}(\varepsilon)-I_{-}(\varepsilon)\right] +\left(\cos{\delta_s}-\cos{\delta_c} \right) \RE{I_{-}(\varepsilon)I^{*}_{+}(\varepsilon)},\\
\label{Resonances-sub-D-1}
\mathcal{D}_{\sigma}^{(1)} (\varepsilon)
&\approx& \frac{1}{2\left(\cos{\delta_s} -\cos{\delta_c}\right)} \left[\sigma \cos{\delta_c}\sin{\delta_s} -\frac{I_{+}(\varepsilon) -I_{-}(\varepsilon)}{2} \left(1 - \cos{\delta_c}\cos{\delta_s} \right)\right],
\end{eqnarray}
\end{widetext}
where
\begin{equation}
\label{Iplusminus}
I_{\pm}(\varepsilon)= \left\langle |u_{\mathbf{k}}(\mathbf{r}_0)|^2  \frac{\Delta(\mathbf{k}) \pm \varepsilon}{\sqrt{|\Delta(\mathbf{k})|^2-\varepsilon^2}} \right\rangle_\varepsilon,
\end{equation}
and we used the definitions in Eqs.~(\ref{Model-ap-def})--(\ref{Model-D-def}).
%(\ref{Model-ah-def}), (\ref{Model-aph-def}), and (\ref{Model-D-def}).
It is straightforward to see that $a_{p} = -a_h^{*} = \left[I_{+}^{*}(\varepsilon)+I_{-}^{*}(\varepsilon)\right]/(4i)$ and $a_{ph} = \left[I_{+}(\varepsilon)-I_{-}(\varepsilon)\right]/(4i)$.

For weak tunneling, the contributions to the denominator which are first order in $|t|^2$ are relevant only for energies approaching the bound states. The energy of these states is obtained from the characteristic equation $\mathcal{D}_{\sigma}^{(0)}(\varepsilon)=0$. The expression for $\mathcal{D}_{\sigma}^{(1)}(\varepsilon)$ was simplified to the form of Eq.~(\ref{Resonances-sub-D-1}) assuming that the equation $\mathcal{D}_{\sigma}^{(0)}(\varepsilon)=0$ has a real-valued positive solution $0\leq \epsilon_{\sigma}\leq \Delta_{\rm min}$. The  spectrum of bound states in the entire energy range $\left[-\Delta_{\rm min}, \Delta_{\rm min}\right]$ follows by  particle-hole symmetry, $\epsilon_{\sigma} =-\epsilon_{-\sigma}$. This justifies restricting our considerations to energies $\varepsilon>0$.

The differential conductance at $\varepsilon<\Delta_{\rm min}$ is determined by Andreev reflections with $|\alpha_{p,h}|=1$ and $|r_{ph,\sigma}|^2+|r_{p,\sigma}|^2=|r_{hp,\sigma}|^2+|r_{h,\sigma}|^2=1$. Then, the conductance (\ref{Resonances-gen-G-def}) reads
\begin{eqnarray}
\label{Resonances-sub-G}
G_{\sigma}(V) &=& \frac{G_n^2}{8 G_Q} \left|I_{+}(eV)+I_{-}(eV)\right|^2 \nonumber\\
&\times&\left[\frac{\theta{\left(V\right)}}{|\mathcal{D}_{\sigma}(eV)|^2} +\frac{\theta{\left(-V\right)}}{|\mathcal{D}_{-\sigma}(-eV)|^2}\right],
\end{eqnarray}
where $G_n = G_{Q} |t|^2$ is the normal-state conductance and $|a_p|^2=|a_h|^2$ for subgap energies.

It is clear from Eq.~(\ref{Resonances-sub-G}) that for subgap energies $G_{\sigma}(V)\sim G_n^2/G_Q$, unless the spectrum contains a bound state and the bias $|eV|$ approaches the bound-state energy $\epsilon_{\sigma}$. The conductance $G_\sigma(V)$ exhibits Lorentzian peaks at $eV=\epsilon_{\sigma}$ with maxima $\sim G_Q$ and peak widths $\sim (G_n/G_Q)|\mathcal{D}_{\sigma}^{(1)}(\epsilon_{\sigma})/[\mathcal{D}_{\sigma}^{(0)}(\epsilon_\sigma)]^\prime |$. Dispensing with the width, we may simplify Eq.~(\ref{Resonances-sub-G}) to
\begin{eqnarray}
\label{Resonances-delta}
G_{\sigma}(V)\! &=& \frac{\pi}{2}W_{\sigma}(|V|) G_n \Delta_{\rm min}
\delta{\left(eV-\epsilon_\sigma\right)}, \\
\label{Resonances-delta-W}
W_{\sigma}(V)\! &=& \frac{1}{4} \left|I_{+}(eV)+I_{-}(eV)\right|^2 \left|\mathcal{D}_{\sigma}^{(1)}(eV)[\mathcal{D}_{\sigma}^{(0)}(eV)]^\prime\right|^{-1}\!\!\!.\nonumber\\
\end{eqnarray}
Accounting for $\propto |t|^2$ corrections to the scattering phases $\delta_s$ and $\delta_c$ would lead to a small shift $\propto G_n/G_Q$ in the resonance energy $\epsilon_{\sigma}$. Below we neglect this small shift and analyze the effects of the gap symmetry on subgap states.

\subsection{Subgap states in an \texorpdfstring{$s$-wave}{s-wave} superconductor}
\label{sec:Resonances-sub-swave}

To set the stage, we consider nodeless gaps that preserve the lattice point and time-reversal symmetries. The simplest example is an $s$-wave gap $\Delta(\mathbf{k}) =\Delta >0$. Then $I_{-}(\varepsilon)I^{*}_{+}(\varepsilon)=1$, as seen from Eq.~(\ref{Iplusminus}). This condition makes $\mathcal{D}_{\sigma}^{(0)}(\varepsilon)$ as defined in  Eq.~(\ref{Resonances-sub-D-0}) insensitive to $\delta_{c}$, which allows us to simplify the characteristic equation $\mathcal{D}_{\sigma}^{(0)}(\varepsilon)=0$ to
\begin{equation}
\label{Resonances-sub-swave-eq}
\cos\delta_s-\sigma\sin\delta_s\frac{\varepsilon}{\sqrt{\Delta^2 -\varepsilon^2}} = 0\,,
\end{equation}
independent of the Bloch functions $u_{\mathbf{k}}(\mathbf{r}_0)$. The solutions of Eq.~(\ref{Resonances-sub-swave-eq}) reproduce the  standard expressions for the energy of YSR states~\cite{Yu:1965,Shiba:1968,Rusinov:1968,Rusinov:1969},
\begin{equation}
\label{Resonances-sub-YSR-eps-swave}
\epsilon_{\sigma} = \sigma \Delta \cos{\delta_{s}}\,,\quad \sigma\cot\delta_s\geq 0.
\end{equation}
Subgap YSR states appear only if $\delta_{s}\neq0$ and are spin polarized, which makes them nondegenerate. The energy of the YSR states is oblivious to the scattering phase $\delta_{c}$, emphasizing the effectiveness of the scattering-theory formulation. One recovers the familiar expression in terms of the coupling strengths $V_{RR}$ and $J_{RR}$ using the relation (\ref{Model-impurity-phase}) \cite{Yu:1965,Shiba:1968,Rusinov:1968,Rusinov:1969},
\begin{equation}
\label{Resonances-sub-YSR-eps-swave-alpha-beta}
\epsilon_\sigma =  \sigma \Delta\frac{1-\alpha^2+\beta^2}{\sqrt{(1-\alpha^2+\beta^2)^2+4\alpha^2}},
\end{equation}
where $\alpha=\pi\nu_0 J_{RR}S$ and $\beta=\pi\nu_0 V_{RR}$. Physically, the spin scattering phase $\delta_s$ depends not only on $J_{RR}$, but also on $V_{RR}$, since the scalar potential modifies the wave-function amplitude at the location of the impurity and hence the strength of spin-dependent scattering.

The YSR states with different spin projections lead to symmetric peaks of equal weight in the differential conductance. Indeed, by using Eqs.~(\ref{Resonances-sub-D-0}), (\ref{Resonances-sub-D-1}), and (\ref{Iplusminus}) in Eqs.~(\ref{Resonances-delta}) and (\ref{Resonances-delta-W}), we obtain the conductance
\begin{eqnarray}
\label{Resonances-sub-swave-G-1}
G_{\sigma}(V) &=& \frac{\pi}{2}W(|V|) G_n \Delta \delta{\left(eV -\epsilon_{\sigma}\right)},\\
\label{Resonances-sub-swave-G-1-W}
W(V) &=&\sqrt{1 - (eV)^2/\Delta^2},
\end{eqnarray}
where the energy of the YSR state is given in Eq.~(\ref{Resonances-sub-YSR-eps-swave}) and is extended to negative values as $\epsilon_{\sigma} =-\epsilon_{-\sigma}$. A nonvanishing conductance in the weak tunneling regime is explained by the resonant Andreev transfer process whereby a particle  virtually tunnels into a YSR state and is retroreflected as a hole. This  resonantly transfers a Cooper pair into the condensate. In view of the particle-hole symmetry of Andreev processes, the corresponding peaks in the differential conductance have heights and symmetric positions. It is worth mentioning, however, that the symmetry of the subgap peaks is a delicate question. Many experimental studies show YSR peaks of different heights in the differential spin-averaged conductance~\cite{Yazdani-Eigler:1997,Franke-Pascual:2011,Ruby-Franke:2016,Thupakula-Massee:2021}. There are also reports of peaks with symmetric heights~\cite{Ruby-Franke:2015}. According to the theoretical analysis in Ref.~\cite{Martin-Mozyrsky:2014,Ruby-Franke:2015}, see also Eqs.~(\ref{Resonances-sub-swave-G-1}) and (\ref{Resonances-sub-swave-G-1-W}), the subgap peaks in the conductance should be particle-hole symmetric and have the same height as long as  inelastic contributions to tunneling are negligible. This was supported in experiment~\cite{Ruby-Franke:2015} by showing that for a normal-metal tip, the peaks become more symmetric as the tip-substrate tunneling increases and Andreev processes become increasingly important.

By using Eqs.~(\ref{Resonances-sub-YSR-eps-swave}), (\ref{Resonances-sub-swave-G-1}), and (\ref{Resonances-sub-swave-G-1-W}), we can clearly see the evolution of the YSR peaks with scattering potential. First, we notice that for a given spin projection $\sigma$, there is only one YSR peak in the conductance. This is explained by the fact that only particles with spin opposite to that of the impurity form a bound state. (We assume  antiferromagnetic exchange between electron and impurity spins.) As the spin scattering phase $\delta_{s}$ increases from $0$ to $\pi/2$, the YSR peak moves from the gap edge $eV=\Delta$ to zero energy, $eV=0$. A further increase of $\delta_{s}$ from $\pi/2$ to $\pi$ makes the peak cross $eV=0$ and reach the other gap edge $eV=-\Delta$ for $\delta_{s}\to \pi$. As follows from particle-hole symmetry, the YSR peak in $G_\sigma(V)$ with the opposite spin projection moves in the opposite direction, i.e., it emerges at $eV=-\Delta$, moves to $eV=0$, and disappears at $eV=\Delta$.

\subsection{Point-symmetry-breaking gaps}
\label{sec:Resonances-sub-point}

YSR states in an $s$-wave superconductor peel off the gap edge at infinitesimally small spin scattering $\delta_s$, and are independent of $\delta_c$, see Eq.~(\ref{Resonances-sub-YSR-eps-swave}). Modulations of the gap $\Delta(\mathbf{k})$ which break the lattice point symmetry make the energy of the YSR states dependent on $\delta_c$. Furthermore, YSR states appear only once the spin scattering phase exceeds a threshold value. As we discuss below, the latter depends on the gap modulation and on $\delta_c$.

To analyze the effect of TRS-preserving gap modulation on YSR states, we write the characteristic equation $\mathcal{D}_{\sigma}^{(0)}(\varepsilon)=0$, see Eq.~(\ref{Resonances-sub-D-0}), in the form
\begin{eqnarray}
\label{exactequation}
&&\cos{\delta_s}+\cos{\delta_c} + \sigma I_{-}(\varepsilon)\sin{\delta_s} \nonumber\\
&&-\left[\sigma \sin{\delta_s} -I_{-}(\varepsilon)\left(\cos{\delta_s}-\cos{\delta_c} \right) \right] I_{+}(\varepsilon)=0.
\end{eqnarray}
One can readily see that the solution of Eq.~(\ref{exactequation}) is independent of $\delta_c$ if $I_{+}(\varepsilon)I_{-}(\varepsilon)=1$. This condition is satisfied at any $\varepsilon$ in an $s$-wave superconductor, yielding Eq.~(\ref{Resonances-sub-YSR-eps-swave}). It is also satisfied for $\varepsilon=0$, regardless of the point symmetry: a zero-energy state appears at $\delta_s=\pm\pi/2$.

The appearance of zero-energy bound states for $\delta_s=\pm\pi/2$ is a universal property of all nodeless TRS-preserving gaps, irrespective of the presence of spatial symmetries. Indeed, according to Bohr-Sommerfeld quantization, a bound state emerges when the total phase accumulated during the Andreev scattering cycle is a multiple of $2\pi$. For a single cycle (see also App.~\ref{sec:App-Model} for a discussion of the scattering processes), one finds the condition $s_{p,\sigma}\alpha_h s_{h,-\sigma}\alpha_p  = 1$ for the formation of the bound state. At $\varepsilon\to0$,  $\alpha_h=\alpha_p=-i$  for any TRS-preserving nodeless gap, see Eq.~(\ref{Model-alpha-def}) for the definition of $\alpha_{p,h}$. In this case, the scattering phase $-\pi/2$ for each of the Andreev retroreflections in the cycle is compensated by the phases acquired in the sequence of particle and hole scatterings off the impurity.

A modulation of the gap, $\Delta_{\rm min}\leq\Delta(\mathbf k)\leq\Delta_{\rm max}$, leads to $I_{+}(\varepsilon)I_{-}(\varepsilon)\neq 1$ and a dependence of the YSR state energy on both $\delta_s$ and $\delta_c$, cf. Eq.~(\ref{exactequation}). We consider the limit of weak modulation, $\Delta_{\rm max}-\Delta_{\rm min}\ll\Delta_{\rm max}+\Delta_{\rm min}$, and dispense with the effect of the Bloch functions in Eq.~(\ref{Iplusminus}) to illustrate the appearance of a threshold value of $\delta_s$ for YSR-state formation. As $\varepsilon\nearrow\Delta_{\rm min}$, the function $I_{+}(\varepsilon)$ diverges logarithmically,
\begin{equation}
\label{lnIplus}
I_{+}(\varepsilon)= 2N\sqrt{\frac{\Delta_{\rm min}}{\Delta^{\prime\prime}_{\rm min}}}\ln{\left(\frac{\Delta_{\rm max}-\Delta_{\rm min}}{\Delta_{\rm min}-\varepsilon}\right)}\,,
\end{equation}
while $I_{-}(\Delta_{\rm min})\sim \sqrt{(\Delta_{\rm max}-\Delta_{\rm min})/(\Delta_{\rm max}+\Delta_{\rm min})}$ and, therefore, $I_{-}(\Delta_{\rm min})\ll 1$. Here $N$ is the number of equivalent minima and $\Delta^{\prime\prime}_{\rm min}\sim (\Delta_{\rm max}-\Delta_{\rm min})$ is the second derivative of the gap along the line of averaging in $\langle\dots\rangle_{\varepsilon}$. The upper cutoff of the logarithmic function in Eq.~(\ref{lnIplus}) is defined up to a factor $\sim 1$. Keeping terms linear in $\delta_s$ in Eq.~(\ref{exactequation}) and implementing the described simplifications in solving it, we find
\begin{eqnarray}
\label{ln-solution}
\ln{\left(\frac{\Delta_{\rm max}-\Delta_{\rm min}}{\Delta_{\rm min}-\epsilon_{\sigma}}\right)}\! &=& \! \frac{\sqrt{{\Delta^{\prime\prime}_{\rm min}/\Delta_{\rm min}}}}{2N} \frac{1+\cos{\delta_c}}{\sigma \delta_s-\delta_s^{\rm th}}, \\
\label{ln-solution-delta-th}
\delta_s^{\rm th}\! &=& \! \left(1-\cos{\delta_c}\right)I_{-}(\Delta_{\rm min})\,.
\end{eqnarray}
This shows that at any $\delta_c\neq 0$, there is a finite threshold $\delta_s^{\rm th}$ for $\delta_s$, at which a YSR level peels off from the quasiparticle continuum. For a nascent YSR state, the difference $\Delta_{\rm min}-\epsilon_{\sigma}$  scales exponentially with $-1/(\delta_s-\delta_s^{\rm th})$. Using Eqs.~(\ref{Resonances-sub-D-0})--(\ref{Resonances-delta-W}) and (\ref{ln-solution}), it is easy to see that the corresponding conductance peak weight scales as $\Delta_{\rm min}-\epsilon_{\sigma}$, and, therefore, is also exponentially small. Accounting for the Bloch functions in Eq.~(\ref{Iplusminus}) does not alter these conclusions.

The threshold value $\delta_s^{\rm th}$ and the value of $\Delta^{\prime\prime}_{\rm min}/\Delta_{\rm min}$ are sensitive to details of the gap function $\Delta(\mathbf k)$. In addition, the range of applicability of Eq.~(\ref{ln-solution}) is severely restricted by the requirement that the logarithmic function therein takes large values.

Away from the gap edges where Eq.~(\ref{ln-solution}) becomes inapplicable, we find that $I_+(\varepsilon)I_-(\varepsilon)-1$ is small for almost any energy. This makes the normalized energies $\epsilon_{\sigma}/\Delta_{\rm min}$ of the YSR levels obtained from the exact characteristic equation (\ref{exactequation}) almost insensitive to the scattering phase $\delta_c$ and allows us to rewrite Eq.~(\ref{exactequation}) as
\begin{equation}
\label{exactequation-app}
2\cos\delta_s-\sigma\sin\delta_s \left[I_{+}(\varepsilon) - I_{-}(\varepsilon)\right] = 0\, ,
\end{equation}
cf., Eq.~(\ref{Resonances-sub-swave-eq}) for $s$-wave gaps. Its low-energy solution is realized at $\delta_s\to \pm \pi/2$ and is similar to that in Eq.~(\ref{Resonances-sub-YSR-eps-swave}), albeit with $\Delta \to \left\langle \Delta(\mathbf k) \right\rangle_0$. A similar simplification of $\mathcal{D}_{\sigma}^{(1)}(\epsilon_{\sigma})$ shows that remarkably, it also becomes independent of $\delta_c$. This allows us to derive a counterpart of Eqs.~(\ref{Resonances-sub-swave-G-1}) and (\ref{Resonances-sub-swave-G-1-W}),
\begin{eqnarray}
\label{Resonances-sub-TRS-G-2}
G_{\sigma}(V) &=& \frac{\pi}{2}W(|V|) G_n\Delta_{\rm min} \delta{\left(eV -\epsilon_{\sigma}\right)}, \\
\label{Resonances-sub-TRS-G-2-W}
W(V) &=& \frac{1}{2\Delta_{\rm min}} \frac{\left[I_{+}(eV)+I_{-}(eV)\right]^2}{\left|I_{+}^\prime(eV)-I_{-}^\prime(eV)\right|}.
\end{eqnarray}

\begin{figure*}[t]
\centering
\subfigure[]{\includegraphics[width=0.45\textwidth]{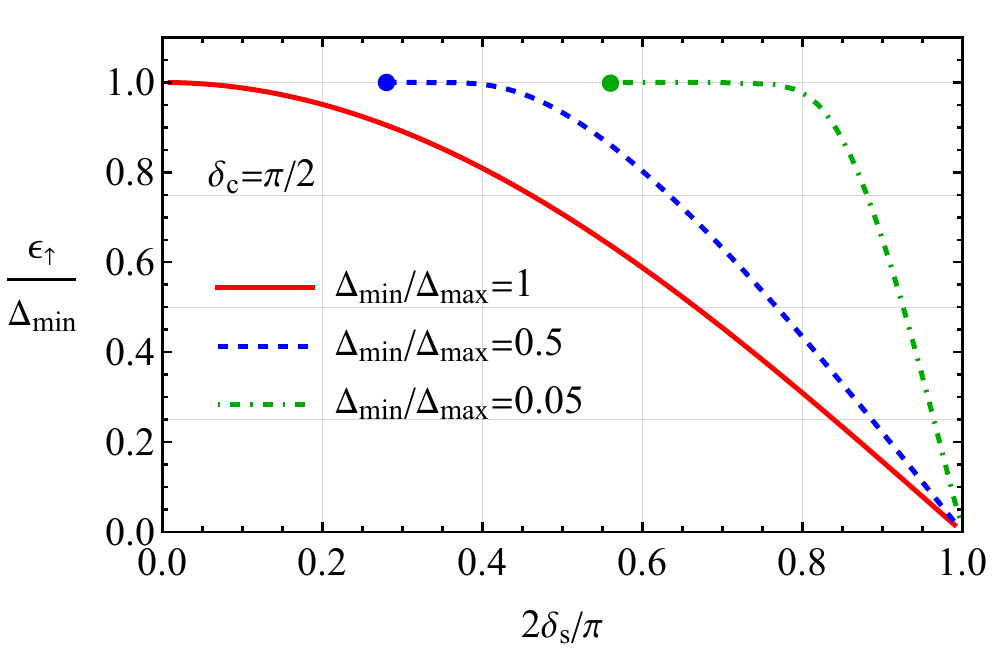}}
\subfigure[]{\includegraphics[width=0.45\textwidth]{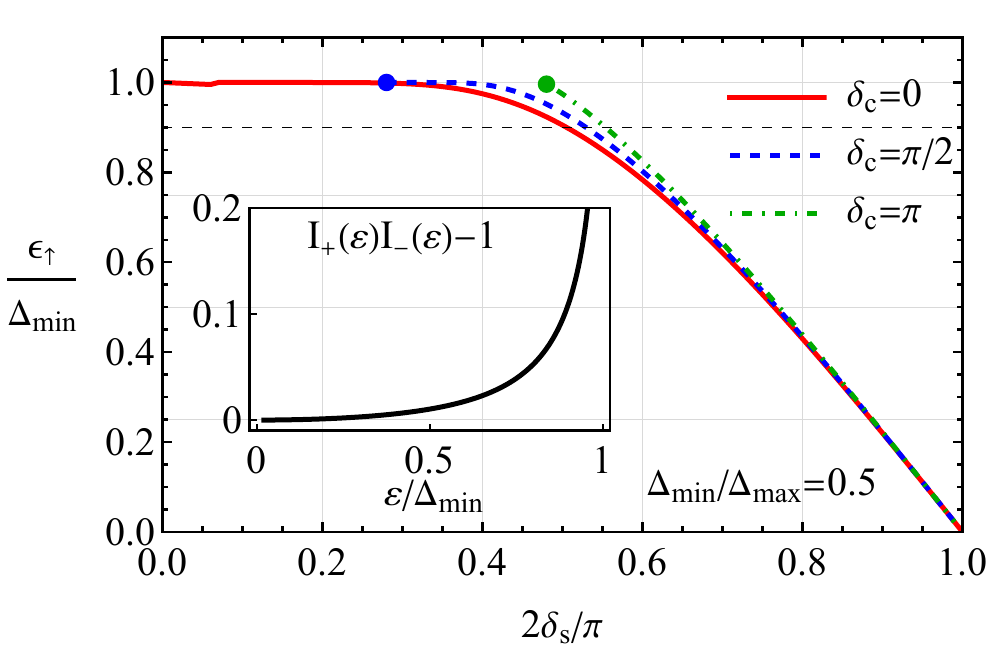}}
\caption{
(a) Normalized energy of subgap states $\epsilon_{\uparrow}/\Delta_{\rm min}$ as a function of $\delta_{s}$ for a few $\Delta_{\rm min}/\Delta_{\rm max}$ at $\delta_{c} = \pi/2$.
(b) Normalized energy of subgap states $\epsilon_{\uparrow}/\Delta_{\rm min}$ as a function of $\delta_{s}$ for a few $\delta_{c}$ at $\Delta_{\rm min}/\Delta_{\rm max}=0.5$. The inset shows that $I_{+}(\varepsilon)I_{-}(\varepsilon)-1$ remains small away from the gap edge. In both panels, we assume a circular Fermi surface (parametrized by the angle $\varphi$), use the $s+d_{x^2-y^2}$ gap $\Delta(\varphi) = \left(\Delta_{\rm max}+\Delta_{\rm min}\right)/2 +\left(\Delta_{\rm max}-\Delta_{\rm min}\right) \cos{(2\varphi)}/2$, and consider tunneling into a high symmetry point. Dots mark the threshold values of the spin scattering phase $\delta_{s}^{\rm th}$ required for the subgap states. The horizontal dashed line in (b) denotes the energy of the subgap state fixed in Fig.~\ref{fig:Resonances-sub-conductance}(a).
}
\label{fig:Resonances-sub-point}
\end{figure*}

To illustrate the appearance of YSR states for gaps breaking the lattice point symmetry, we show numerical results for an $s+d_{x^2-y^2}$ superconductor in Fig.~\ref{fig:Resonances-sub-point}. In agreement with Eqs.~(\ref{ln-solution}) and (\ref{ln-solution-delta-th}), there is a threshold value for the scattering phase $\delta_s$ when $\delta_{c}\neq0$. The threshold is larger for more anisotropic gaps. The weak sensitivity to the scattering phase $\delta_c$ in a broad energy region excluding near-edge energies is evident from Fig.~\ref{fig:Resonances-sub-point}(b), where the inset shows $I_{+}(\varepsilon)I_{-}(\varepsilon)-1$. The smallness of the latter explains the insensitivity to $\delta_c$ away from the gap edges.

\begin{figure*}[t]
\centering
\subfigure[]{\includegraphics[width=0.45\textwidth]{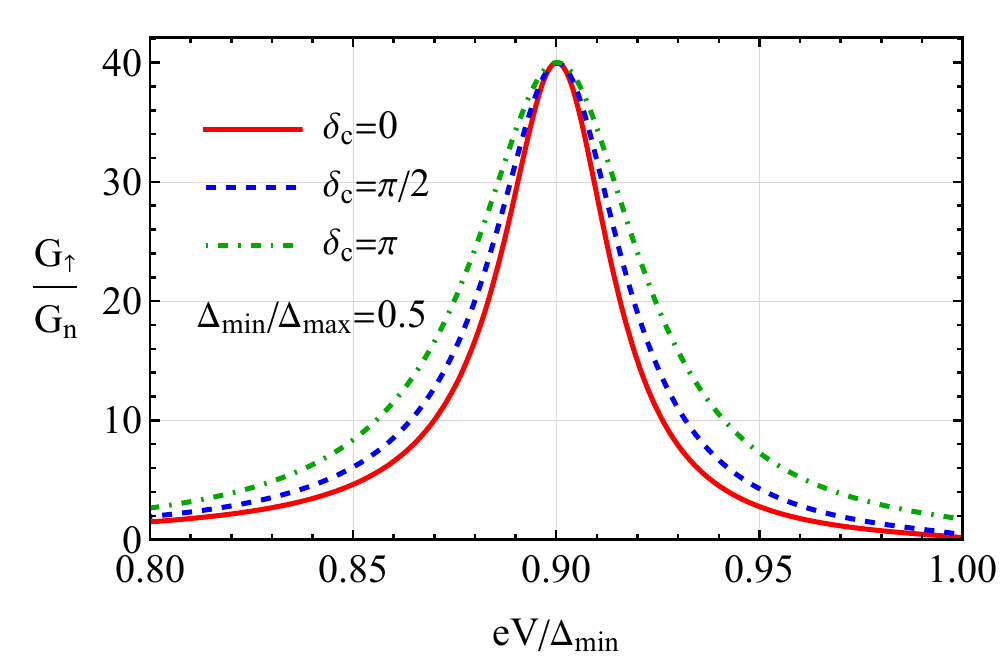}}
\subfigure[]{\includegraphics[width=0.45\textwidth]{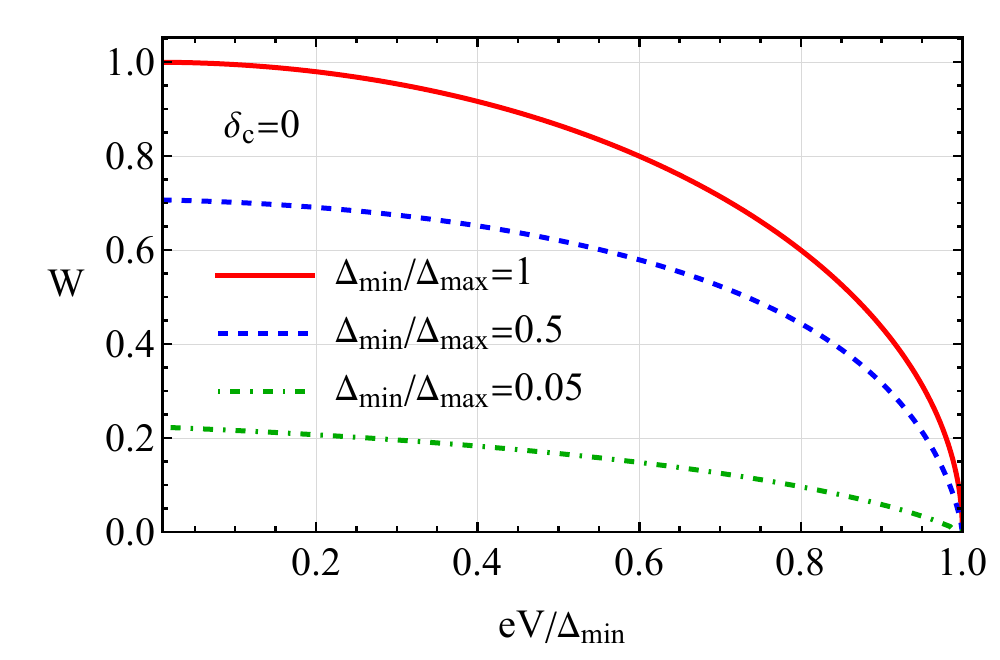}}
\caption{
(a) Normalized conductance as a function of bias voltage for several values of $\delta_{c}$ at $\epsilon_{\uparrow}=0.9\,\Delta_{\rm min}$ as indicated by the horizontal dashed line in Fig.~\ref{fig:Resonances-sub-point}(b).
(b) Weight of the peaks in the conductance $W$, see Eqs.~(\ref{Resonances-delta-W}), (\ref{Resonances-sub-swave-G-1-W}), and (\ref{Resonances-sub-TRS-G-2-W}), as a function of bias voltage for several values of $\Delta_{\rm min}/\Delta_{\rm max}$. In all panels, we assume a circular Fermi surface, consider tunneling into a high symmetry point, and use the $s+d_{x^2-y^2}$ gap $\Delta(\varphi) = \left(\Delta_{\rm max}+\Delta_{\rm min}\right)/2 +\left(\Delta_{\rm max}-\Delta_{\rm min}\right) \cos{(2\varphi)}/2$; we fix $G_{n}/G_{Q} = 0.05$ in panel (a) and $\delta_c=0$ in panel (b).
}
\label{fig:Resonances-sub-conductance}
\end{figure*}

In STS experiments, the bound states are manifested as peaks of the differential conductance. Figure~\ref{fig:Resonances-sub-conductance}(a) illustrates this for an impurity creating a state with energy $\epsilon_{\uparrow}=0.9\,\Delta_{\rm min}$. Even for such close proximity to the gap edge, the dependence on the scattering phase $\delta_c$ is weak; it becomes undetectable for energies $\epsilon_{\uparrow}\lesssim 0.6\,\Delta_{\rm min}$. The anisotropy of the gap has a much larger effect on the conductance, which is evident from the weight of the peaks shown in Fig.~\ref{fig:Resonances-sub-conductance}(b). While the overall shape remains parabolic-like, the weight decreases for anisotropic gaps. Indeed, since the scattering phase $\delta_c$ affects only a narrow near-edge region, in the rest of the energy range the weight resembles the result in Eqs.~(\ref{Resonances-sub-swave-G-1}) and (\ref{Resonances-sub-swave-G-1-W}) with $\Delta \to \left\langle \Delta(\mathbf k) \right\rangle_0$.

\subsection{TRS-broken gaps}
\label{sec:Resonances-sub-TRS}

To investigate the effects of broken TRS in the characteristic equation for bound states, we return to Eq.~(\ref{Resonances-sub-D-0}) with $I_{\pm}(\varepsilon)$ defined in Eq.~(\ref{Iplusminus}). Due to particle-hole symmetry ($\epsilon_\sigma=-\epsilon_{-\sigma}$), we can still constrain the solutions of $\mathcal{D}_{\sigma}^{(0)}(\varepsilon)=0$ to the interval $[0,\Delta_{\rm min}]$. In a superconductor with broken TRS, even a potential scatterer ($\delta_s=0$) may nucleate a subgap state. Unlike the YSR level, such a state would be spin degenerate, $\epsilon_\uparrow=\epsilon_\downarrow$. The introduction of spin-dependent scattering would split it. At small splitting, both nondegenerate levels may fall into the subgap interval.

These results follow from Eq.~(\ref{Resonances-sub-D-0}). At $\delta_s=0$, $\mathcal{D}_{\sigma}^{(0)}(\varepsilon)=0$ takes the form
\begin{equation}
\label{exactequation-nodeltas}
1+\tan^2{\left(\frac{\delta_c}{2}\right)} \RE{I_{-}(\varepsilon)I^{*}_{+}(\varepsilon)}=0\,.
\end{equation}
As in the TRS-preserving case, $I_{+}(\varepsilon)$ diverges at $\varepsilon\nearrow\Delta_{\rm min}$. This divergence is logarithmic for an anisotropic gap ($\Delta_{\rm max}\neq\Delta_{\rm min}$), and carries over to a divergence of $\RE{I_{-}(\varepsilon)I^{*}_{+}(\varepsilon)}$,
\begin{widetext}
\begin{equation}
\label{ReIplusminus}
\RE{I_{-}(\varepsilon)I^{*}_{+}(\varepsilon)} = A\{\Delta({\mathbf k)}\} \ln{\left(\frac{\Delta_{\rm max}-\Delta_{\rm min}}{\Delta_{\rm min}-\varepsilon}\right)}\,,\,\,\,
A\{\Delta({\mathbf k)}\} = \sum_{j} |u_{\mathbf{k}_j}(\mathbf{r}_0)|^2 \sqrt{\frac{\Delta_{\rm min}}{|\Delta_j|^{\prime \prime}}}  \RE{I_{-}(\Delta_{\rm min})\left(1 +\frac{\Delta_{j}^{*}}{\Delta_{\rm min}}\right)},
\end{equation}
\end{widetext}
where $j$ labels equivalent gap minima. Here, the absolute value of the coefficient  $|A\{\Delta({\mathbf k)}\}|\sim 1$, but its sign is sensitive to the  structure of the gap $\Delta({\mathbf k})$. (We further elaborate on this at the end of this section). If $A\{\Delta({\mathbf k)}\}<0$, a bound state (albeit an exponentially shallow one) appears at arbitrarily weak potential scattering $\delta_c\ll1$,
\begin{equation}
\label{ln-solution-noTRS}
\ln{\left(\frac{\Delta_{\rm max}-\Delta_{\rm min}}{\Delta_{\rm min}-\epsilon_{\sigma}}\right)} = -\frac{1}{A\{\Delta({\mathbf k})\}}\frac{4}{\delta_c^2}\,.
\end{equation}
Similar to the previous section, the weight of the conductance peak associated with this bound state scales as $\Delta_{\rm min}-\epsilon_{\sigma}$, and therefore is also exponentially small.

In the absence of both, TRS and lattice point symmetry, there is no general rule controlling the sign of $A\{\Delta({\mathbf k})\}$, so that a weak potential scatterer may or may not lead to a subgap state. If point symmetry is preserved and $\Delta({\mathbf k})$ belongs to its nontrivial representation, then there is at least one bound state for any scattering. To demonstrate this, we first note that the terms in $I_{\pm}(\varepsilon)$ stemming from $\Delta({\mathbf k})$ in the numerators of the respective expressions average to zero, see Eq.~(\ref{Iplusminus}). As a result,
\begin{equation}
\label{Iplusminus-symm}
I_{+}(\varepsilon)=-I_{-}(\varepsilon)=I_0(\varepsilon)
\equiv\varepsilon\left\langle  \frac{|u_{\mathbf{k}}(\mathbf{r}_0)|^2}{\sqrt{|\Delta(\mathbf{k})|^2-\varepsilon^2}} \right\rangle_\varepsilon.
\end{equation}
Returning to the notations $\delta_\uparrow$ and $\delta_\downarrow$, we may rewrite the characteristic equation $\mathcal{D}_{\sigma}^{(0)}(\varepsilon)=0$ in the form
\begin{equation}
\label{pointsymm-no-TRS}
\left[\cos\delta_\uparrow-\sigma\sin\delta_\uparrow \, I_0(\varepsilon)\right] \left[\cos\delta_\downarrow+\sigma\sin\delta_\downarrow \, I_0(\varepsilon)\right]=0,
\end{equation}
demonstrating the existence of bound states for any scattering. In the two special cases $\delta_s=0$ and $\delta_c=0$, the state (with $\varepsilon>0$) is doubly degenerate. The factorized form of Eq.~(\ref{pointsymm-no-TRS}) has a symmetry origin: scattering off the impurity occurs only \emph{every other} cycle of Andreev reflection, which allows a particle (or a hole) return to the zero angular momentum state. At fixed spin, there are no more than two positive-energy solutions of Eq.~(\ref{pointsymm-no-TRS}), $\epsilon^p_\sigma$ and $\epsilon^h_\sigma$.

The eigenfunction of a spin-up ($\sigma=\uparrow$) solution $\epsilon^p_\uparrow$ obtained from setting the first factor in Eq.~(\ref{pointsymm-no-TRS}) to zero contains an $s$-wave component of a spin-up particle. The eigenfunction of a spin-up solution $\epsilon^h_\uparrow$ obtained from setting to zero the second factor in Eq.~(\ref{pointsymm-no-TRS}) contains an $s$-wave component of a spin-down hole. The symmetry of the bound-state eigenfunctions makes them invisible in the maximally-symmetric STM setting, which we consider here. To detect these states, tunneling should occur at a point off the impurity site. Even then, the spin selection rules may prevent their observation in a tunneling experiment. For example, if Eq.~(\ref{pointsymm-no-TRS}) has two spin-up solutions, only one will show up in $G_\uparrow(V)$. A weak violation of the point group symmetry would lift the selection rule, while weakly affecting the energies of the localized states.

\begin{figure*}[!ht]
\centering
\subfigure[]{\includegraphics[width=0.45\textwidth]{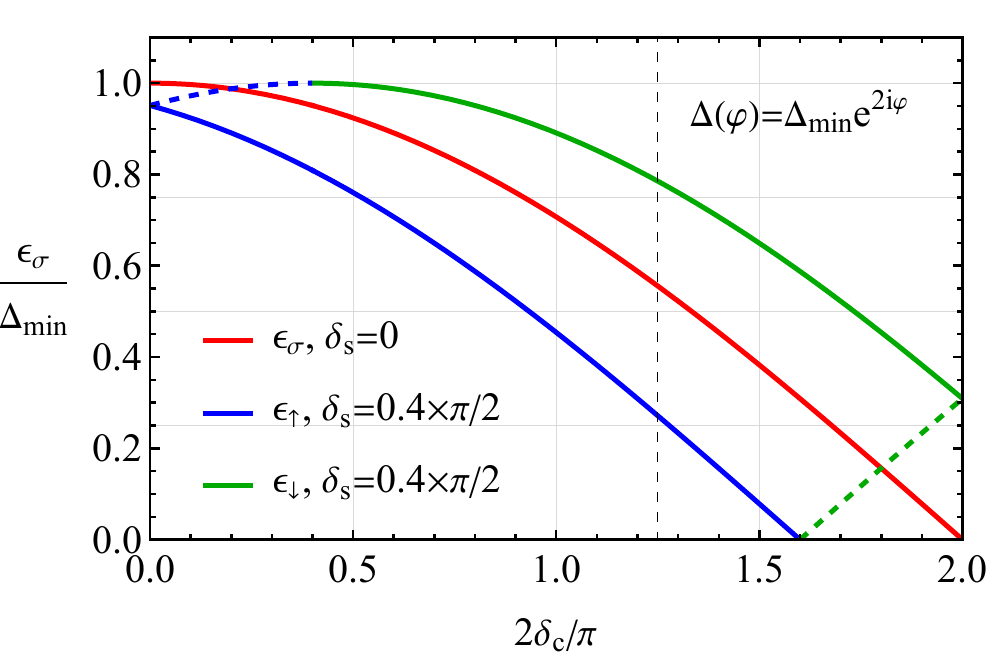}}
\subfigure[]{\includegraphics[width=0.45\textwidth]{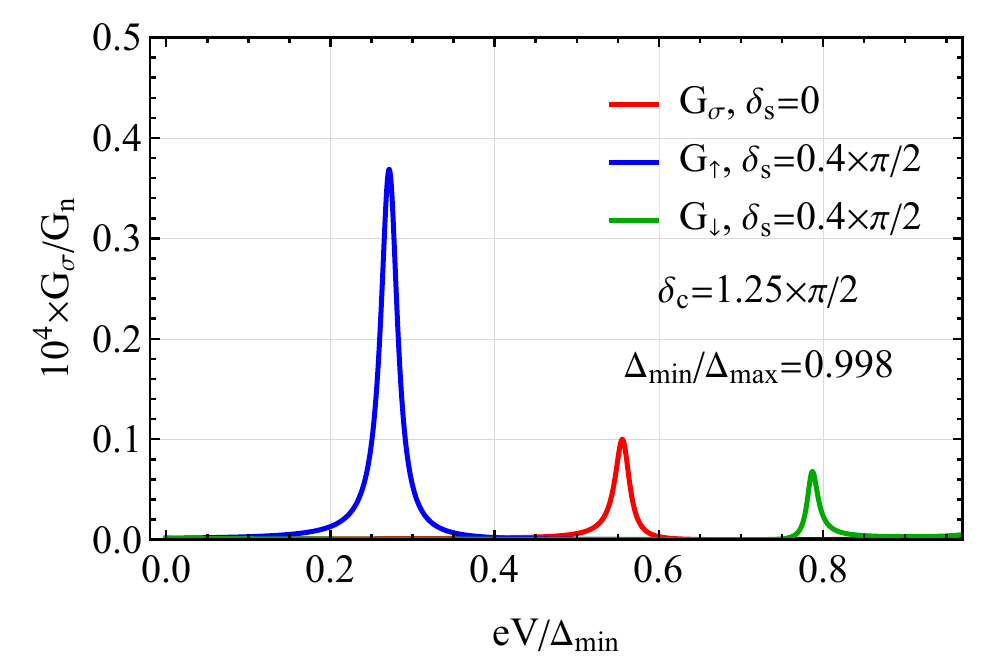}}
\caption{
(a) Normalized energy of the subgap states $\epsilon_{\sigma}/\Delta_{\rm min}$ as a function of $\delta_{c}$ for $\delta_{s}=0$ (solid red line) and $\delta_{s}=0.4\,\pi/2$ (blue and green lines). Spin-dependent scattering splits the spin-degenerate level (red line) into two states $\epsilon_{\uparrow}$ (blue line) and  $\epsilon_{\downarrow}$ (green line). Dashed lines denote solutions of Eq.~(\ref{Iplusminus-symm}) with the same spin polarization, see also Appendix~\ref{sec:App-d+id} for the exact dispersion relation.
(b) Normalized conductance as a function of bias voltage for several values of $\delta_{s}$. We fix $\delta_c=1.25\,\pi/2$, see the vertical dashed line in panel (a), and $G_{n}/G_{Q} = 0.05$. We assume a circular Fermi surface and tunneling into a high symmetry point, use the $s+d_{x^2-y^2}+id_{xy}$ gap $\Delta(\varphi) = \left(\Delta_{\rm max}+\Delta_{\rm min}\right)/2 +\left(\Delta_{\rm max}-\Delta_{\rm min}\right) e^{2i\varphi}/2$, and fix $\Delta_{\rm min}/\Delta_{\rm max}=1$ in panel (a) and $\Delta_{\rm min}/\Delta_{\rm max}=0.998$ in panel (b).
}
\label{fig:Resonances-sub-TRS}
\end{figure*}

Comparing Eq.~(\ref{exactequation-nodeltas}) with the full equation, $\mathcal{D}_{\sigma}^{(0)}(\varepsilon)=0$ [cf. Eqs.~(\ref{Resonances-sub-D-0}) and (\ref{Iplusminus})], we see that the splitting of a spin-degenerate level is linear in $\delta_s$. We illustrate this effect in Fig.~\ref{fig:Resonances-sub-TRS} for the $d_{x^2-y^2}+id_{xy}$ gap $\Delta(\mathbf{k}) = \Delta_{\rm min} e^{2i\varphi}$; see also Appendix~\ref{sec:App-d+id} for the exact dispersion relation. Notice that since the differential conductance for a $d_{x^2-y^2}+id_{xy}$ gap vanishes for symmetry reasons, we added a small admixture of an $s$-wave component breaking point symmetry in Fig.~\ref{fig:Resonances-sub-TRS}(b). The weight of the peaks becomes larger at smaller energies, which agrees with the behavior in Fig.~\ref{fig:Resonances-sub-conductance}(b) for TRS-preserving gaps.

\subsection{Topology and bound states}
\label{sec:Resonances-sub-topology}

Before ending our discussion of bound states within a nodeless gap, we comment on the interplay between a superconductor's topology and bound states. A celebrated consequence of a material's nontrivial topology is the bulk-boundary correspondence dictating the presence of edge states at interfaces with topologically trivial materials. It was conjectured in Ref.~\cite{Slager-Balents:2015} that an impurity necessarily creates a bound state in a topological TRS-preserving superconductor.

A straightforward application of our formalism allows us to check whether this conjecture carries over to TRS-breaking superconductors. The answer is in general negative: Depending on the momentum dependence of the gap $\Delta(\mathbf{k})$, a 2D topological superconductor may or may not carry impurity-induced bound states. Below we provide two respective examples.

In a superconductor with $\Delta(\mathbf{k}) = \Delta_0 + \Delta_1 e^{2i\varphi}$ (exemplifying an $s+d_{x^2-y^2}+id_{xy}$ gap), the sign of $A\{\Delta({\mathbf k)}\}$ in Eq.~(\ref{ln-solution-noTRS}) and therefore the presence of the subgap states correlate with a nontrivial topological invariant $|\mathcal{C}|$. Here, the topological invariant equals $|\mathcal{C}|=2$ if $|\Delta_0|<|\Delta_1|$ and $\mathcal{C}=0$ if $|\Delta_0|>|\Delta_1|$; see, e.g., Refs.~\cite{Volovik:1997,Volovik:book-2003}. This correlation is, however, nonuniversal.

To demonstrate this, we consider the gap function
\begin{equation}
\label{Resonances-sub-topology-Delta-def}
\Delta(\varphi) = \Delta_0 e^{-a \cos{\varphi}} e^{i\left(\varphi - \sin{\varphi}\right)},
\end{equation}
which is topologically nontrivial, but need not support bound states. Here, $\Delta_0$ is a dimensionful parameter, while $a>0$ is a dimensionless parameter that determines the minimal value of $|\Delta(\varphi)|$, i.e., $\Delta_{\rm min} =\Delta_0 e^{-a}$; at $a\to\infty$, $\Delta_{\rm min} \to 0$.

It is straightforward to check that the Chern number for the gap in Eq.~(\ref{Resonances-sub-topology-Delta-def}) is nontrivial for any value of $a$. Indeed, assuming the BCS approximation, the Chern number reads
\begin{eqnarray}
\label{Resonances-sub-topology-C-def}
\mathcal{C} &=& \int_0^{2\pi}\frac{d\varphi}{4\pi} \int_{-\infty}^{\infty}d\xi \frac{\Delta_0^2 e^{-2a \cos{\varphi}} \left(\cos{\varphi}-1\right)}{\left(\Delta_0^2 e^{-2a \cos{\varphi}} +\xi^2\right)^{3/2}} \nonumber\\
&=& \int_0^{2\pi}\frac{d\varphi}{2\pi} \left(\cos{\varphi}-1\right) =-1,
\end{eqnarray}
see, e.g., Ref.~\cite{Volovik:1997} for details.

For a gap that has a single type of minima $\Delta_{\rm min}$, the criterion of the appearance of bound states discussed in Sec.~\ref{sec:Resonances-sub-TRS}, see Eqs.~(\ref{ReIplusminus}) and (\ref{ln-solution-noTRS}), reduces to the sign of $A\{\Delta({\mathbf k)}\}$, $\sign{A\{\Delta({\mathbf k)}\}} = \sign{\RE{I_{-}(\Delta_{\rm min})}}$. Bound states exist if $\RE{I_{-}(\Delta_{\rm min})}<0$. By using Eqs.~(\ref{Iplusminus}) and (\ref{Resonances-sub-topology-Delta-def}), we obtain
\begin{equation}
\label{Resonances-sub-topology-I-def}
\RE{I_{-}(\Delta_{\rm min})} = \int_0^{2\pi}\! \frac{d\varphi}{2\pi}\frac{e^{-a\cos{\varphi}}\cos{\left(\varphi - \sin{\varphi}\right)} -e^{-a}}{\sqrt{e^{-2a\cos{\varphi}}-e^{-2a}}}.
\end{equation}
Here, for simplicity, we assumed a circular Fermi surface and tunneling into a high-symmetry point with $|u_{\mathbf{k}}(\mathbf{r}_0)|=1$. At $a\to0$, the sign of $\RE{I_{-}(\Delta_{\rm min})}$ is determined by the numerator, $\cos{\left(\varphi - \sin{\varphi}\right)} -1 \leq 0$. Therefore, bound states are allowed for $a\to0$. In the opposite limit, $a\to\infty$, we have $\lim_{a\to\infty} \RE{I_{-}(\Delta_{\rm min})} = 2\pi J_1(1)>0$, where $J_n(x)$ is the Bessel function of the first kind. Since $\RE{I_{-}(\Delta_{\rm min})}$ as a function of $a$ changes sign only once at $a^{\star}\approx0.9$, bound states disappear at $a>a^{\star}$. Thus, the gap function (\ref{Resonances-sub-topology-Delta-def}) provides a counterexample supporting our statement that there is no general relation between the Chern number of the gap and the number of subgap states.

\section{Resonances in nodal-gap superconductors}
\label{sec:Resonances}

We can also apply our scattering approach to investigate impurity-induced resonances near the Fermi level in nodal-gap superconductors. Such resonances remain narrow as long as their energy is small, $\varepsilon\ll\Delta_{\rm max}$. The energies of the resonances, of course, satisfy the particle-hole-symmetry condition. However, the strength of the resonances as manifested in the differential-conductance peak amplitudes develops strong asymmetry. The scattering theory, along with the condition $\varepsilon\ll\Delta_{\rm max}$, allows us to provide an intuitive interpretation of this effect. The asymmetry of the tunneling DOS persists to higher energies~\cite{Balatsky-Zhu:rev-2006}, albeit its description becomes more involved.

For a nodal gap, the denominator of the scattering amplitudes $\mathcal{D}_{\sigma}(\varepsilon)$ is a complex-valued function of energy. The poles of the scattering amplitudes, determined by the characteristic equation $\mathcal{D}_{\sigma}^{(0)}(\varepsilon)=0$, shift away from the real axis, implying a finite width of the conductance peaks. This allows us to take the limit $|t|\to 0$ in the full expression for  $\mathcal{D}_{\sigma}(\varepsilon)$ and replace it with $\mathcal{D}_{\sigma}^{(0)}(\varepsilon)$. At energies $\varepsilon\ll\Delta_{\rm max}$, the quasiparticle DOS remains small, leading a parametrically small ${\rm Im}\,\mathcal{D}_{\sigma}^{(0)}(\varepsilon)$ compared to ${\rm Re}\,\mathcal{D}_{\sigma}^{(0)}(\varepsilon)$, apart from the  vicinity of the poles. Therefore, we can evaluate $\mathcal{D}_{\sigma}^{(0)}(\varepsilon)$ iteratively. We first determine the position of the resonances by solving ${\rm Re}\,\mathcal{D}_{\sigma}^{(0)}(\varepsilon)=0$. We subsequently expand ${\rm Re}\,\mathcal{D}_{\sigma}^{(0)}(\varepsilon)$ around the solutions ($\varepsilon=\epsilon_\sigma$) and evaluate ${\rm Im}\,\mathcal{D}_{\sigma}^{(0)}(\epsilon_\sigma)$. The latter allows us to determine the width $\Gamma_{\sigma}$ of the resonances.

In performing the first task, we may still use Eq.~(\ref{Resonances-sub-D-0}), after judiciously taking the limit $\varepsilon\to 0$. The resulting expression for ${\rm Re}\,\mathcal{D}_{\sigma}^{(0)}(\varepsilon)$ depends crucially on the presence of lattice point-group symmetry. In its absence, the characteristic equation takes the form
\begin{eqnarray}
\label{nodal-ReD}
&&\cos{\delta_s}+\cos{\delta_c} +\left(\cos{\delta_s}-\cos{\delta_c} \right)\! \left|\left\langle\! |u_{\mathbf{k}}(\mathbf{r}_0)|^2  \frac{\Delta(\mathbf{k})}{|\Delta(\mathbf{k})|}\!\right\rangle_0\right|^2 \nonumber\\
&&- 2\sigma \sin{\delta_s} \left\langle |u_{\mathbf{k}}(\mathbf{r}_0)|^2  \frac{\varepsilon}{|\Delta(\mathbf{k})|}\right\rangle_{\varepsilon\leq |\Delta(\mathbf{k})|} =0\,.
\end{eqnarray}
Here the term $|\langle\dots\rangle_0|^2\neq 0$ is allowed by the broken symmetry. The average $\langle\dots\rangle$ in the last term is performed only over the part of the Fermi line where $|\Delta(\mathbf{k})|\geq\varepsilon$, see Appendix~\ref{sec:app-small-bias} for details. Because of the nodal gap structure, this average scales with energy as $\langle\dots\rangle \propto \varepsilon \ln{(\Delta_{\rm max}/\varepsilon)}$. For Eq.~(\ref{nodal-ReD}) to have a solution at $\varepsilon\ll\Delta_{\rm max}$, spin scattering must be present, $\sin\delta_s\neq 0$. In addition, the scattering phases must satisfy the stringent condition
\begin{equation}
\label{stringent}
\frac{\cos{\delta_c}}{\cos{\delta_s}} \approx \frac{\left|\left\langle |u_{\mathbf{k}}(\mathbf{r}_0)|^2  \Delta(\mathbf{k})/|\Delta(\mathbf{k})|\right\rangle_0\right|^2+1}{\left|\left\langle |u_{\mathbf{k}}(\mathbf{r}_0)|^2  \Delta(\mathbf{k})/|\Delta(\mathbf{k})|\right\rangle_0\right|^2-1}\,.
\end{equation}
The left- and right-hand sides depend, respectively, on the properties of the impurity and the host material. It is thus difficult to expect this relation to hold without fine-tuning. Therefore, we conclude that, in the absence of point-group symmetry, the appearance of low-energy resonances is improbable. This prompts us to focus on the symmetric case in the remainder of this section.

If lattice point symmetry is preserved and $\Delta({\mathbf k})$ belongs to its nontrivial representation, then $\left\langle |u_{\mathbf{k}}(\mathbf{r}_0)|^2  \Delta(\mathbf{k})/|\Delta(\mathbf{k})|\right\rangle_0=0$, and the corresponding characteristic equation takes a form similar to Eq.~(\ref{pointsymm-no-TRS}),
\if 0
\begin{widetext}
\begin{equation}
\label{pointsymm-nodal}
\left[\cos\delta_\uparrow-\sigma\sin\delta_\uparrow \left\langle |u_{\mathbf{k}}(\mathbf{r}_0)|^2 \frac{\varepsilon}{|\Delta(\mathbf{k})|}\right\rangle_{\varepsilon\leq |\Delta(\mathbf{k})|}
\right] \left[\cos\delta_\downarrow+\sigma\sin\delta_\downarrow  \left\langle |u_{\mathbf{k}}(\mathbf{r}_0)|^2 \frac{\varepsilon}{|\Delta(\mathbf{k})|}\right\rangle_{\varepsilon\leq |\Delta(\mathbf{k})|}
\right] = 0.
\end{equation}
\end{widetext}
\fi
\begin{eqnarray}
&&\left[\cos\delta_\uparrow-\sigma\sin\delta_\uparrow \left\langle |u_{\mathbf{k}}(\mathbf{r}_0)|^2 \frac{\varepsilon}{|\Delta(\mathbf{k})|}\right\rangle_{\varepsilon\leq |\Delta(\mathbf{k})|}
\right] \label{pointsymm-nodal} \\
&&\times\left[\cos\delta_\downarrow+\sigma\sin\delta_\downarrow  \left\langle |u_{\mathbf{k}}(\mathbf{r}_0)|^2 \frac{\varepsilon}{|\Delta(\mathbf{k})|}\right\rangle_{\varepsilon\leq |\Delta(\mathbf{k})|}
\right] = 0. \nonumber
\end{eqnarray}
Upon evaluation of the averages $\langle\dots\rangle$ with logarithmic accuracy, see Appendix~\ref{sec:app-small-bias}, we find the positive-energy solutions
%\clearpage
\begin{eqnarray}
\label{resonances}
&&\epsilon_{\sigma, \eta} =\frac{\varepsilon^{\star}\sigma \eta \cot\delta_\eta}{\ln{\left|\Delta_{\rm max}/(\varepsilon^{\star}\cot{\delta_\eta})\right|}}\,,\,\, \varepsilon^{\star} \! = \frac{\pi}{\sum_{j} |u_{\mathbf{k}_j}(\mathbf{r}_0)|^2/|\Delta^{\prime}_j|}, \nonumber \\ &&\frac{\varepsilon^{\star}|\cot{\delta_\eta}|}{\Delta_{\rm max}}\ll 1\,,\quad \sigma \eta \cot\delta_\eta\geq 0\,.
\end{eqnarray}
Here $\eta=\uparrow,\downarrow$, the derivative $\Delta^{\prime}_j$ is taken over a dimensionless vector tangential to the Fermi line, and $j$ labels the gap nodes. As before, particle-hole symmetry $\epsilon_{\sigma, \eta} =- \epsilon_{-\sigma, \eta}$ allowed us to focus on solutions in the interval $\left[0,\Delta_{\rm max}\right]$. If the last condition in Eq.~(\ref{resonances}) is not fulfilled, then the bound state is located in the interval $\left[-\Delta_{\rm max},0\right]$. For a given direction of spin $\sigma$, there are at most two solutions of Eq.~(\ref{pointsymm-nodal}), see the last condition in Eq. (\ref{resonances}). The typical value of the energy scale introduced in Eq.~(\ref{resonances}) is $\varepsilon^\star\sim\Delta_{\rm max}$. Therefore, for a resonance to be close to the Fermi level, the corresponding scattering phase shift should satisfy the condition $|\delta_\eta-\pi/2|\ll 1$.

Expanding the denominator $\mathcal{D}_{\sigma}^{(0)}(\varepsilon)$ in the vicinity of the resonance $\varepsilon=\epsilon_{\sigma,\eta}$, it is straightforward to obtain the width of the resonance $\Gamma_{\sigma, \eta} = \mbox{Im}\,\mathcal{D}_{\sigma}^{(0)}(\epsilon_{\sigma,\eta})/\mbox{Re}\,\left[\mathcal{D}_{\sigma}^{(0)}(\epsilon_{\sigma,\eta})\right]^{\prime}$,
\begin{equation}
\label{Resonances-inter-res-Gamma}
\Gamma_{\sigma,\eta} = \frac{\pi}{2} \frac{|\epsilon_{\sigma, \eta}|}{\ln{\left|\Delta_{\rm max}/\epsilon_{\sigma, \eta}\right|}}.
\end{equation}
Here, the imaginary part of the average is free of divergence, $\mbox{Im}\,\left\langle |u_{\mathbf{k}}(\mathbf{r}_0)|^2\,\varepsilon/|\Delta(\mathbf{k})|\right\rangle_{\varepsilon\geq |\Delta(\mathbf{k})|} = \pi \varepsilon/(2 \varepsilon^{\star})$. As one can see, $\Gamma_{\sigma, \eta}\ll\epsilon_{\sigma,\eta}$ for $\delta_\eta\to\pi/2$ leading to well-defined low-energy resonances in the strong-scattering regime. We note in passing that the condition $\delta_\eta\to\pi/2$ corresponds to resonant scattering in the absence of superconductivity, as well. However, in the normal state, the resonances are typically broad (with the width substantially exceeding $\Delta_{\rm max}$)~\cite{Hirschfeld-Einzel:1988}.

As mentioned at the beginning of this section, the energies of the impurity resonances are particle-hole symmetric. This symmetry, however, is not manifested in the differential conductance. The reason is due to a crucial difference between the conductance $G_{\sigma}(V)$ determined by the full scattering amplitudes, which depend on both energies and wave functions of the resonant states, and the particle-hole-symmetric density of energy levels determined by the structure of the poles of these amplitudes. To show this, we evaluate the conductance assuming strong scattering, point-symmetry preserving gaps, and tunneling into high-symmetry points. Expanding the numerator and denominator of the conductance in the vicinity of the resonance and within logarithmic accuracy, we obtain
\begin{equation}
\label{resonances-G}
G_{\sigma}(V) = \frac{\pi G_n}{2\ln^2|\Delta_{\rm max}/\epsilon_{\sigma,\sigma}|} \frac{|eV| \varepsilon^{\star}}{\left(eV-\epsilon_{\sigma,\sigma}\right)^2+\Gamma_{\sigma,\sigma}^2},
\end{equation}
see Eqs.~(\ref{Resonances-gen-G-def}) and (\ref{Resonances-gen-rp2}) for the conductance expressed in terms of scattering amplitudes as well as Eqs.~(\ref{resonances}) and (\ref{Resonances-inter-res-Gamma}) for the definitions of $\epsilon_{\sigma,\sigma}$ and $\Gamma_{\sigma,\sigma}$. In the conductance (\ref{resonances-G}), the energies of the bound states given in Eq.~(\ref{resonances}) were extended to negative values by  particle-hole symmetry, $\epsilon_{\sigma, \eta} =- \epsilon_{-\sigma, \eta}$. As one can see from Eq.~(\ref{resonances-G}), one of the two poles of the energy denominator $\mathcal{D}_{\sigma}^{(0)}(\varepsilon)$, cf. Eqs.~(\ref{pointsymm-nodal}) and (\ref{resonances}), was canceled by the numerator representing the eigenfunction of the state. This demonstrates the crucial role of the resonance-state structure.

\begin{figure*}[t]
\centering
\subfigure[]{\includegraphics[width=0.45\textwidth]{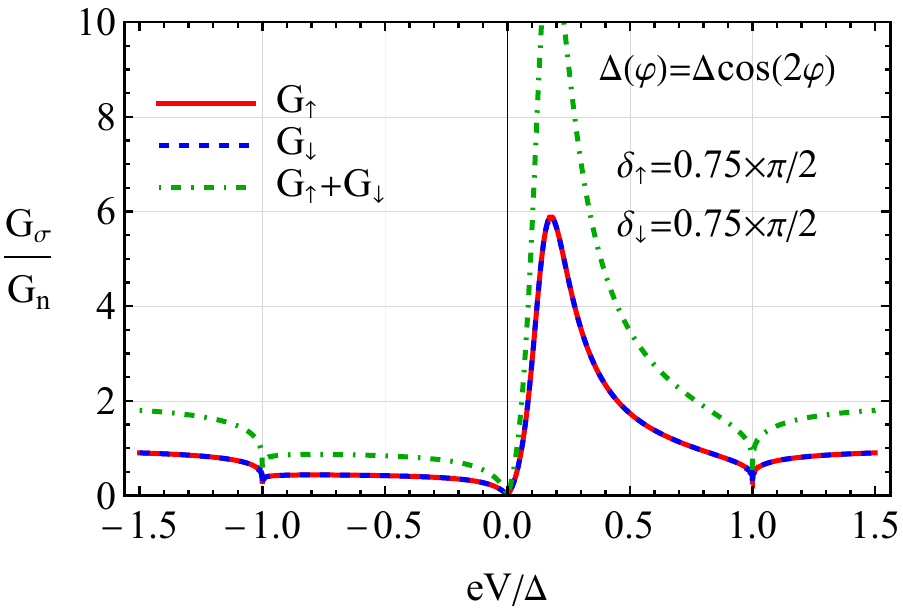}}
\hspace{0.01\textwidth}
\subfigure[]{\includegraphics[width=0.45\textwidth]{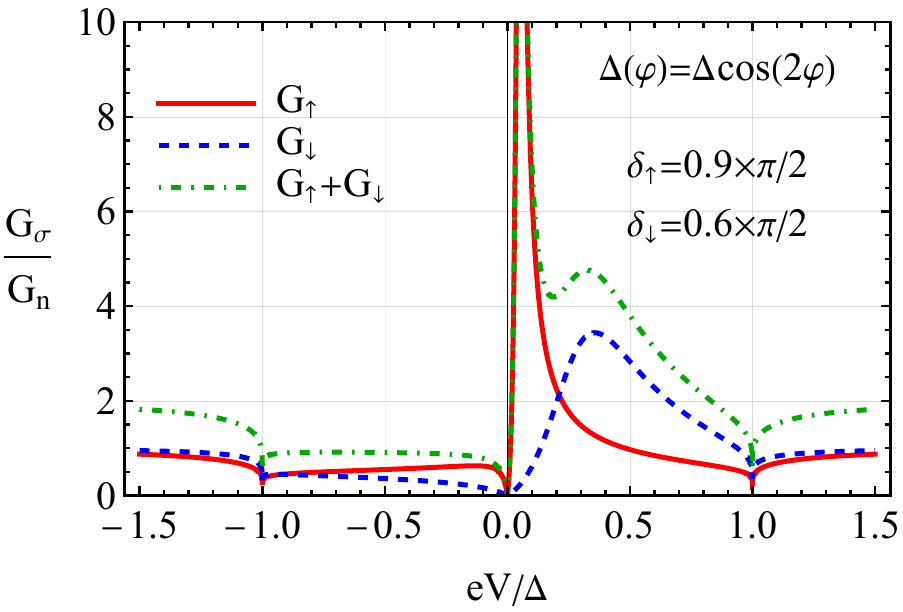}}
\caption{
Differential conductance as a function of voltage bias for several values of the scattering phases. (a) Spin-independent scattering $\delta_{\uparrow} = \delta_{\downarrow} = 0.75\, \pi/2$ leads to a single spin-polarized peak in $(G_{\uparrow}+ G_{\downarrow})/G_n$. (b) The peak is split by adding a spin-dependent scattering phase with $\delta_{\uparrow} = 0.9\, \pi/2$ and $\delta_{\downarrow} = 0.6\, \pi/2$. Solid red and dashed blue lines correspond to spin-up ($G_{\uparrow}/G_n$) and spin-down ($G_{\downarrow}/G_n$) conductances, respectively. Dash-dotted green lines denote the conductance summed over spin projections $(G_{\uparrow}+ G_{\downarrow})/G_n$.
In both panels, we assume a circular Fermi surface, use the $d$-wave gap $\Delta(\mathbf{k}) = \Delta \cos{(2\varphi)}$, and consider tunneling into a high symmetry point.
}
\label{fig:Resonances-inter-d-wave-nonmag}
\end{figure*}

The conductance in Eq.~(\ref{resonances-G}) is manifestly particle-hole asymmetric, with the strongest asymmetry achieved for potential scattering $\delta_{\uparrow} = \delta_{\downarrow}$. There is a single spin-degenerate peak in the conductance in this case, as one may see from Eq.~(\ref{resonances-G}). Spin-dependent scattering lifts the degeneracy and splits the peaks with different spin polarizations, ultimately leading to two symmetric peaks at $\pi/2-\delta_{\uparrow} = \delta_{\downarrow}-\pi/2$. We illustrate the splitting of the peaks in Fig.~\ref{fig:Resonances-inter-d-wave-nonmag} for the $d$-wave gap $\Delta(\mathbf{k}) = \Delta \cos{(2\varphi)}$.

We end this section by emphasizing the different nature of the subgap conductance in superconductors with nodeless and nodal gaps. In the former case, an electron incoming from the STM tip is reflected as a hole at $|eV|\leq\Delta_{\rm min}$. This process is inevitably particle-hole symmetric, resulting in the symmetry of the differential conductance, $G_\sigma(V)=G_{-\sigma}(-V)$, see Sec.~\ref{sec:Resonances-sub-swave}. On the contrary, electron tunneling in a nodal-gap superconductor turns into (to the first order in $G_n$) a quasiparticle propagating into the superconductor. In this case, the symmetry of the conductance may be violated, but only at biases $|eV|\leq\Delta_{\rm max}$, see also the Supplemental Material in Ref.~\cite{Sukhachov-Glazman:2022-STM} for the detailed calculations and Appendix~\ref{sec:App-Resonances-above} for the results at $|eV|\geq\Delta_{\rm max}$. The origin of this symmetry violation is in the interference between processes responsible for the conversion of an impinging particle in the tip into a quasiparticle propagating in the superconductor. There are direct conversion processes and processes in which the particle completes a few Andreev scattering cycles before propagating into the superconductor as a quasiparticle. Both types of processes are allowed for nodal gaps.

\section{Summary}
\label{sec:Summary}

In this paper, we studied the role of the gap and Bloch-function symmetries in the formation and properties of impurity-induced states in a 2D superconductor, and investigated manifestations of these states in scanning tunneling spectra. We applied the scattering approach developed by us in Ref.~\cite{Sukhachov-Glazman:2022-STM} to treat weak tunneling between the STM tip and the superconductor, extending it here to arbitrarily strong scattering off an impurity in the superconductor. We consider both spin-independent and spin-dependent scattering. The framework is conveniently formulated in terms of the phase shifts of electron scattering off the impurity and is summarized in Sec.~\ref{sec:Model}, see Eqs.~(\ref{Model-sp}) and (\ref{Model-rph})--(\ref{Resonances-gen-G-def}).
Our work provides an intuitive approach to calculating the spectrum of impurity-induced states and the differential conductance in an STM setting. In addition, it complements and clarifies previous studies of impurity resonances.

In agreement with the literature, the bound states for nodeless gaps that preserve TRS and lattice point symmetry are found only for spin-dependent scattering and are independent of the potential scattering phase $\delta_c$, see Secs.~\ref{sec:Resonances-sub-swave} as well as Eqs.~(\ref{Resonances-sub-YSR-eps-swave}) and (\ref{Resonances-sub-swave-G-1}). These YSR states are particle-hole symmetric, spin-polarized, and lead to peaks of the same weight in the differential conductance summed over both spin projections, see also Eq.~(\ref{Resonances-sub-swave-G-1}) and the subsequent discussion.

The modulation of a nodeless gap resulting from the breaking of the lattice point symmetry makes YSR states dependent on the potential scattering phase $\delta_c$. In this case, a YSR state appears near the gap edge once the spin scattering phase exceeds a threshold value determined by $\delta_c$ and the anisotropy of the gap, see Eqs.~(\ref{ln-solution}) and (\ref{ln-solution-delta-th}) as well as Fig.~\ref{fig:Resonances-sub-point}. Away from the narrow energy range near the gap edges, the bound state becomes almost insensitive to the scattering phase $\delta_c$ and resembles the YSR state for an $s$-wave gap. In STS experiments, the weight of the peaks of the differential conductance decreases toward the gap edges, where it becomes sensitive to $\delta_c$, see Eqs.~(\ref{Resonances-sub-TRS-G-2}) and (\ref{Resonances-sub-TRS-G-2-W}), and Fig.~\ref{fig:Resonances-sub-conductance}.

Breaking of TRS in a superconductor makes potential scattering capable of creating spin-degenerate subgap states, see Sec.~\ref{sec:Resonances-sub-TRS} and Eq.~(\ref{ln-solution-noTRS}). Introducing spin-dependent scattering lifts the degeneracy, leading to two spin-polarized states, as is illustrated in Fig.~\ref{fig:Resonances-sub-TRS}. At small splitting, both states may fall into the subgap interval and result in two peaks of different weights in the differential conductance. The predicted bound states can be probed by spin-polarized STM/STS~\cite{Schneider-Wiesendanger:2020,Wang-Wiesendanger:2020}. While the corresponding superconducting gaps may be topological, as we discussed in Sec.~\ref{sec:Resonances-sub-topology}, bound states for pointlike impurities generically do not enjoy topological protection.

For nodal gaps, well-pronounced resonances appear at low energies for strong scattering potentials of any type if the gap preserves lattice point symmetry, see Eqs.~(\ref{resonances}) and (\ref{Resonances-inter-res-Gamma}). The observation of low-energy resonances in the absence of point-group symmetry might require fine-tuning, see Eq.~(\ref{stringent}). The resonance states are spin degenerate (spin polarized) for potential (spin-dependent) scattering and always respect particle-hole symmetry, see Eq.~(\ref{resonances}). Contrary to the density of energy levels, the differential conductance (\ref{resonances-G}), determined by the full scattering amplitudes rather than their poles, is particle-hole asymmetric. The strongest asymmetry is found for potential scattering leading to a single spin-degenerate peak in the conductance. Spin-dependent scattering lifts the degeneracy and splits the peaks with different spin polarizations as is illustrated in Fig.~\ref{fig:Resonances-inter-d-wave-nonmag}.

The theory presented in this work combines the general scattering formalism with the eikonal approximation for quasiparticles scattering off an impurity in a superconductor. It offers a convenient and intuitive scheme for investigating the impurity-induced resonances in the excitation spectrum of superconductors. In this work, we exemplified the theory on single-band superconductors with nontrivial scalar gap functions. This provides the foundation for considering materials with multiple bands and multicomponent superconducting order parameters, which would extend its applicability to iron pnictides~\cite{Ng-Avishai:2009,Hirschfeld-Mazin:2011,Beaird-Zhu:2012} and twisted bilayer graphene~\cite{Lake-Senthil:2022}.

\begin{acknowledgments}
We are grateful to Alexander Balatsky, Eduardo H. da Silva Neto, and Pavel Volkov for discussions and comments. The work was supported by NSF Grant No.~DMR-2002275 and by the Office of Naval Research (ONR) under Award No.~N00014-22-1-2764 at Yale University, and by Deutsche Forschungsgemeinschaft through CRC 183 and a joint ANR-DFG project (TWISTGRAPH) at Freie Universit\"{a}t Berlin. P.O.S. acknowledges support through the Yale Prize Postdoctoral Fellowship in Condensed Matter Theory. L.I.G. thanks Freie Universit\"{a}t Berlin for hosting his stay as a CRC 183 Mercator fellow.
\end{acknowledgments}

\appendix

\section{Details of the scattering theory}
\label{sec:App-Model}

\subsection{Scattering cycles}
\label{sec:App-ST-cycles}

To derive the scattering amplitudes given in Eqs.~(\ref{Model-rph}) and (\ref{Model-rp}), we use the same approach as for a spin-unpolarized STM tip developed in Ref.~\cite{Sukhachov-Glazman:2022-STM}. The model of the contact between  STM tip and 2D superconductor as well as key approximations are outlined in Sec.~\ref{sec:Model}. In this appendix, we focus on the technical details of the derivation.

The tip emits the particle wave $\psi_{p,\mathbf{k},\sigma}^{(1)} = t_{p,\sigma} u_{\mathbf{k}}(\mathbf{r}_0)$ into a 2D superconductor, where $u_{\mathbf{k}}({\bf r}_0)$ is the Bloch function at the tunneling point $\mathbf{r}_0$ and $t_{p,\sigma}$ is the transmission matrix element. We assume that the tip is spin polarized along or opposite to the direction of the spin of the impurity and denote the projection of the spin of impinging particles as $\sigma = \uparrow, \downarrow$.
The gap anisotropy becomes imprinted in the retroreflected hole wave,
\begin{equation}
\label{App-ST-cycles-psi-h}
\psi_{h,\mathbf{k},-\sigma}=\alpha_p(\mathbf{k},\varepsilon)\psi_{p,\mathbf{k},\sigma}^{(1)}.
\end{equation}
Here, the Andreev retroreflection amplitudes for particles $\alpha_p(\mathbf{k},\varepsilon)$ and holes $\alpha_h(\mathbf{k},\varepsilon)$ are
\begin{equation}
\label{Model-alpha-def}
\alpha_{p,h}(\mathbf{k},\varepsilon)=\exp{\left\{\pm i\arg{\left[\frac{\varepsilon}{\Delta(\mathbf{k})}\right]}-i\arccos{\left|\frac{\varepsilon}{\Delta(\mathbf{k})}\right|}\right\}}.
\end{equation}
The analytical continuation to $\varepsilon >\left|\Delta(\mathbf{k})\right|$ is determined by the requirement $|\alpha_{p,h}|\leq 1$ and the sign $+(-)$ corresponds to the particle-to-hole (hole-to-particle) conversion. In addition, we assume spin-singlet superconductors with $\Delta(\mathbf{k})$ being a scalar.

Only part of the wave (\ref{App-ST-cycles-psi-h}), i.e., $\hat{P}\psi_{h,\mathbf{k},-\sigma}$, can interact with the tip, while the other part, $\left(\hat{I}-\hat{P}\right) \psi_{h,\mathbf{k},-\sigma}$ is oblivious to its presence. Here, the single-channel nature of the electron scattering of an incoming state is effected by the projection operator
\begin{equation}
\label{App-ST-cycles-Phat}
\hat{P} \psi^{\rm in}_{\mathbf{k},\sigma} = u_{\mathbf{k}}({\bf r}_0)\!\!\!  \sum_{\xi({\mathbf{k}}^\prime)=\varepsilon} \!\!\! u_{\mathbf{k}^\prime}^*({\bf r}_0)\psi^{\rm in}_{\mathbf{k}^\prime,\sigma}\! \equiv\! u_{\mathbf{k}}({\bf r}_0)\langle u_{\mathbf{k}^\prime}^*({\bf r}_0)\psi^{\rm in}_{\mathbf{k}^\prime,\sigma}\rangle_\varepsilon;
\end{equation}
the unit operator is denoted by $\hat{I}$. The averaging is performed over the Fermi contour $\xi(\mathbf{k})=\varepsilon$ with $\xi(\mathbf{k})$ being the energy dispersion relations measured from the Fermi energy; see Eq.~(\ref{Model-average-def}) for the explicit definition.
Therefore, the hole wave scattered off the tip is
\begin{equation}
\label{App-ST-cycles-psi-h-1}
\psi_{h,\mathbf{k},-\sigma}^{(1)} =\left[\hat{I}-\left(1-s_{h,-\sigma}\right)\hat{P}\right]\psi_{h,\mathbf{k},-\sigma},
\end{equation}
where $s_{h,-\sigma}$ is the scattering matrix element for holes associated with a single impurity in the absence of the tip.

Within the 2D material, the hole wave $\psi_{h,\mathbf{k},-\sigma}^{(1)}$ is retroreflected as a particle wave $\alpha_{h}(\mathbf{k},\varepsilon)\psi_{h,\mathbf{k},-\sigma}^{(1)}$. After interacting with the tip, the scattered particle wave is
\begin{eqnarray}
\label{App-ST-cycles-psi-p-2}
\psi_{p,\mathbf{k},\sigma}^{(2)} &=& \left[\hat{I}-\left(1-s_{p,\sigma}\right)\hat{P}\right]\alpha_{h}(\mathbf{k},\varepsilon) \psi_{h,\mathbf{k},-\sigma}^{(1)} \nonumber\\
&=&\left[\hat{I}-\left(1-s_{p,\sigma}\right)\hat{P}\right] \alpha_{h}(\mathbf{k},\varepsilon) \left[\hat{I}-\left(1-s_{h,-\sigma}\right)\hat{P}\right] \nonumber\\
&\times&\alpha_p(\mathbf{k},\varepsilon)  t_{p, \sigma} u_{\mathbf{k}}(\mathbf{r}_0).
\end{eqnarray}
The particle wave $\psi_{p,\mathbf{k},\sigma}^{(2)}$ can be again retroreflected as a hole wave $\alpha_{p}(\mathbf{k},\varepsilon)\psi_{p,\mathbf{k},\sigma}^{(2)}$. This closes a single Andreev scattering cycle in which the impinging particle interacts with the tip and is retroreflected as a hole. For particles to be retroreflected as particles, an additional half cycle should be added.

By summing over  cycles, we derive the following scattered hole and particle wave functions:
\begin{eqnarray}
\label{App-ST-cycles-psi-h-out}
\psi_{h,\mathbf{k},-\sigma}^{\rm out}\!\! &=&\! \sum_{n=0}^{\infty} \hat{L}_{\sigma}^{n} \alpha_p(\mathbf{k},\varepsilon)t_{p,\sigma} u_{\mathbf{k}}(\mathbf{r}_0),\\
\label{App-ST-cycles-psi-p-out}
\psi_{p,\mathbf{k},\sigma}^{\rm out}\!\! &=&\! \alpha_h(\mathbf{k},\varepsilon) \left[\hat{I}-\left(1-s_{h,-\sigma}\right)\hat{P}\right] \psi_{h,\mathbf{k},-\sigma}^{\rm out},
\end{eqnarray}
where the Andreev cycle operator is defined as
\begin{eqnarray}
\label{App-ST-cycles-L-def}
\hat{L}_{\sigma} &=& \alpha_p(\mathbf{k},\varepsilon) \left[\hat{I}-\left(1-s_{p,\sigma}\right)\hat{P}\right] \nonumber\\
&\times&\alpha_h(\mathbf{k},\varepsilon) \left[\hat{I}-\left(1-s_{h,-\sigma}\right)\hat{P}\right].
\end{eqnarray}

Finally, by using the outgoing wave functions (\ref{App-ST-cycles-psi-h-out}) and (\ref{App-ST-cycles-psi-p-out}), we derive the Andreev and normal scattering amplitudes for particles with the spin projection $\sigma$:
\begin{widetext}
\begin{eqnarray}
\label{App-ST-cycles-rph-fin}
r_{ph,\sigma} &=& t_{h, -\sigma}^{\prime} t_{p, \sigma} \left\langle u_{\mathbf{k}}^{*}({\bf r}_0)\sum_{n=0}^\infty \hat{L}_{\sigma}^n\, \alpha_{p}(\mathbf{k},\varepsilon)u_{\mathbf{k}}({\bf r}_0)\right\rangle_\varepsilon
\equiv t_{h, -\sigma}^{\prime} t_{p, \sigma} \left\langle u_{\mathbf{k}}^{*}({\bf r}_0)M_{\sigma}(\mathbf{k})\right\rangle_\varepsilon, \\
\label{App-ST-cycles-rp-fin}
r_{p,\sigma} &=& s_{p,\sigma}^{\prime} + t_{p, \sigma}^{\prime} t_{p, \sigma} \left\langle u_{\mathbf{k}}^{*}({\bf r}_0) \alpha_h(\mathbf{k},\varepsilon) \left[\hat{I}-\left(1-s_{h,-\sigma}\right)\hat{P}\right] M_{\sigma}(\mathbf{k})\right\rangle_\varepsilon,
\end{eqnarray}
\end{widetext}
where the normal reflection amplitude includes also a scattering matrix element in the tip $s_{p,\sigma}^{\prime}$.

\subsection{Scattering amplitudes}
\label{sec:App-ST-cycles-Andreev}

Let us perform the summation over all scattering cycles in Eqs.~(\ref{App-ST-cycles-rph-fin}) and (\ref{App-ST-cycles-rp-fin}), and derive the expressions for the scattering amplitudes $r_{ph,\sigma}$ and $r_{p,\sigma}$. Symbolically, we have the following relation in Eq.~(\ref{App-ST-cycles-rph-fin}):
\begin{equation}
\label{App-ST-cycles-Andreev-Mphi}
M_{\sigma}(\mathbf{k})=(\hat{I}-\hat{L}_{\sigma})^{-1}\alpha_p(\mathbf{k},\varepsilon)u_{\mathbf{k}}({\bf r}_0),
\end{equation}
which can be rewritten as an integral equation
\begin{equation}
\label{App-ST-cycles-Andreev-M-int}
\left(\hat{I} -\hat{L}_{\sigma}\right)M_{\sigma}(\mathbf{k}) = \alpha_p(\mathbf{k},\varepsilon) u_{\mathbf{k}}(\mathbf{r}_0).
\end{equation}
Since the operator $\hat{L}_{\sigma}$ defined in Eq.~(\ref{App-ST-cycles-L-def}) has a separable kernel, the integral equation (\ref{App-ST-cycles-Andreev-M-int}) can be brought to a set of algebraic equations; see, e.g., Ref.~\cite{Morse-Feshbach:book} for a generic approach. Indeed, the action of the operator $\hat{L}_{\sigma}$ on $M_{\sigma}(\mathbf{k})$ reads
\begin{eqnarray}
\label{App-ST-cycles-Andreev-M-int-expl}
\hat{L}_{\sigma} M_{\sigma}(\mathbf{k}) &=&
\alpha_{p}(\mathbf{k},\varepsilon)\alpha_{h}(\mathbf{k},\varepsilon) M_{\sigma}(\mathbf{k}) \nonumber\\
&-&\left(1-s_{h,-\sigma}\right) \alpha_{p}(\mathbf{k},\varepsilon) \alpha_{h}(\mathbf{k},\varepsilon)u_{\mathbf{k}}(\mathbf{r}_0) M_1 \nonumber\\
&-& \left(1-s_{p,\sigma}\right) \alpha_{p}(\mathbf{k},\varepsilon) u_{\mathbf{k}}(\mathbf{r}_0) M_2 \nonumber\\
&+& \left(1-s_{p,\sigma}\right)\left(1-s_{h,-\sigma}\right) \alpha_{p}(\mathbf{k},\varepsilon) u_{\mathbf{k}}(\mathbf{r}_0) \nonumber\\
&\times&\left\langle |u_{\mathbf{k}}(\mathbf{r}_0)|^2 \alpha_{h}(\mathbf{k},\varepsilon) \right\rangle_{\varepsilon} M_1,
\end{eqnarray}
where
\begin{equation}
\label{App-ST-cycles-Andreev-M1-M2}
M_1 = \left\langle u_{\mathbf{k}}^{*}(\mathbf{r}_0) M_{\sigma}(\mathbf{k})\right\rangle_{\varepsilon}, \,\,
M_2 =\left\langle u_{\mathbf{k}}^{*}(\mathbf{r}_0) \alpha_h(\mathbf{k},\varepsilon) M_{\sigma}(\mathbf{k})\right\rangle_{\varepsilon}.
\end{equation}

By substituting Eq.~(\ref{App-ST-cycles-Andreev-M-int-expl}) into Eq.~(\ref{App-ST-cycles-Andreev-M-int}), we find the following expression for $M_{\sigma}(\mathbf{k})$ in terms of $M_1$ and $M_2$:
\begin{widetext}
\begin{eqnarray}
\label{App-ST-cycles-Andreev-M-eq}
M_{\sigma}(\mathbf{k}) &=& \frac{u_{\mathbf{k}}(\mathbf{r}_0) \alpha_p(\mathbf{k},\varepsilon)}{1-\alpha_{p}(\mathbf{k},\varepsilon)\alpha_h(\mathbf{k},\varepsilon)} + \left(1 - s_{p,\sigma}\right)\left(1 - s_{h,-\sigma}\right) \frac{u_{\mathbf{k}}(\mathbf{r}_0) \alpha_p(\mathbf{k},\varepsilon)}{1-\alpha_p(\mathbf{k},\varepsilon)\alpha_h(\mathbf{k},\varepsilon)} \left\langle |u_{\mathbf{k}}(\mathbf{r}_0)|^2 \alpha_{h}(\mathbf{k},\varepsilon) \right\rangle_{\varepsilon} M_1
\nonumber\\
&-& \left(1-s_{h,-\sigma}\right) \frac{u_{\mathbf{k}}(\mathbf{r}_0) \alpha_p(\mathbf{k}) \alpha_h(\mathbf{k},\varepsilon)}{1-\alpha_p(\mathbf{k}) \alpha_h(\mathbf{k},\varepsilon)} M_1 -\left(1-s_{p,\sigma}\right) \frac{u_{\mathbf{k}}(\mathbf{r}_0) \alpha_p(\mathbf{k},\varepsilon)}{1-\alpha_p(\mathbf{k},\varepsilon)\alpha_h(\mathbf{k},\varepsilon)} M_2.
\end{eqnarray}
\end{widetext}
Then, by using this result in Eq.~(\ref{App-ST-cycles-Andreev-M1-M2}), we obtain the set of algebraic equations for $M_1$ and $M_2$:
\begin{widetext}
\begin{eqnarray}
\label{App-ST-cycles-Andreev-M-eq-1}
&&\left[1 -\left(1 - s_{p,\sigma}\right)\left(1 - s_{h,-\sigma}\right) a_{p} \left\langle |u_{\mathbf{k}}(\mathbf{r}_0)|^2 \alpha_{h}(\mathbf{k},\varepsilon) \right\rangle_{\varepsilon} +\left(1 - s_{h,-\sigma}\right) a_{ph}\right]M_1 +\left(1 - s_{p,\sigma}\right) a_{p} M_2 =a_{p},\\
\label{App-ST-cycles-Andreev-M-eq-2}
&&\left(1 - s_{h,-\sigma}\right) \left\{a_h - \left[1 + \left(1 - s_{p,\sigma}\right)a_{ph}\right] \left\langle |u_{\mathbf{k}}(\mathbf{r}_0)|^2 \alpha_{h}(\mathbf{k},\varepsilon) \right\rangle_{\varepsilon}
\right\}M_1 +\left[1 +\left(1 - s_{p,\sigma}\right) a_{ph}\right]M_2=a_{ph}.
\end{eqnarray}
\end{widetext}
Here, we used shorthand notations $a_{p}$, $a_{h}$, and $a_{ph}$ given in Eqs.~(\ref{Model-ap-def}), (\ref{Model-ah-def}), and (\ref{Model-aph-def}), respectively. By solving the system of algebraic equations (\ref{App-ST-cycles-Andreev-M-eq-1}) and (\ref{App-ST-cycles-Andreev-M-eq-2}), we obtain
\begin{widetext}
\begin{eqnarray}
\label{App-ST-cycles-Andreev-M-eq-sol-1}
M_1 &=& \frac{a_{p}}{1 + \left(2-s_{p,\sigma} -s_{h,-\sigma}\right)a_{ph} +\left(1 -s_{p,\sigma}\right) \left(1 -s_{h,-\sigma}\right) \left(a_{ph}^2 -a_{p} a_{h}\right)},\\
\label{App-ST-cycles-Andreev-M-eq-sol-2}
M_2 &=& \frac{a_{ph} +\left(1 -s_{h,-\sigma}\right)\left[a_{ph}^2 -a_p\left(a_{h} -\left\langle |u_{\mathbf{k}}(\mathbf{r}_0)|^2 \alpha_{h}(\mathbf{k},\varepsilon) \right\rangle_{\varepsilon}\right) \right]}{1 + \left(2-s_{p,\sigma}-s_{h,-\sigma}\right)a_{ph} +\left(1 -s_{p,\sigma}\right)\left(1 -s_{h,-\sigma}\right) \left(a_{ph}^2 -a_{p} a_{h}\right)}.
\end{eqnarray}
\end{widetext}
The above equations allow us to derive the Andreev and normal reflection amplitudes
%\clearpage
\begin{eqnarray}
\label{App-ST-cycles-Andreev-rph-fin}
r_{ph,\sigma} &=& t_{h,-\sigma}^{\prime} t_{p,\sigma} M_1,\\
\label{App-ST-cycles-Andreev-rp-fin}
r_{p,\sigma} &=& s_{p,\sigma}^{\prime} + t_{p,\sigma}^{\prime} t_{p,\sigma}\Big[M_2 -\left(1-s_{h,-\sigma}\right) \nonumber\\
&\times&\left\langle |u_{\mathbf{k}}(\mathbf{r}_0)|^2 \alpha_h(\mathbf{k},\varepsilon)\right\rangle_{\varepsilon} M_1\Big].
\end{eqnarray}
The final expressions are given in Eqs.~(\ref{Model-rph}) and (\ref{Model-rp}).

\section{Low-energy properties of averages}
\label{sec:app-small-bias}

In this section, we discuss the averages in the coefficients $a_{p,h}$ and $a_{ph}$ defined in Eqs.~(\ref{Model-ap-def}), (\ref{Model-ah-def}), and (\ref{Model-aph-def}) at small energies $\varepsilon\ll \Delta_{\rm max}$. The case of nodeless gaps is straightforward and is covered in the main text. Therefore, we consider small-energy expansion for nodal gaps. Due to the presence of the nodes, the coefficients $a_{p,h}$ and $a_{ph}+1/2$ have real and imaginary parts even in the presence of TRS. The corresponding approximate expressions can be estimated by linearizing $\Delta(\mathbf{k})$ in the vicinity of the nodes $\Delta_j=0$ and performing the integration over the small intervals of the Fermi line in the vicinity of the nodes at $\varepsilon \geq |\Delta(\mathbf{k})|$ (imaginary part) or away from the nodes at $\varepsilon \leq |\Delta(\mathbf{k})|$ (real part).

We find the leading asymptote for the integral in $a_{ph}+1/2$ at $\varepsilon\geq |\Delta(\mathbf{k})|$:
\begin{eqnarray}
\label{app-SB-2-Ag}
\left\langle \frac{|u_{\mathbf{k}}(\mathbf{r}_0)|^2 \,\varepsilon}{\sqrt{|\Delta(\mathbf{k})|^2-\varepsilon^2}} \right\rangle_{\varepsilon\geq |\Delta(\mathbf{k})|}\!\!\! &\approx&\!\! i\left\langle |u_{\mathbf{k}}(\mathbf{r}_0)|^2 \frac{\varepsilon}{|\Delta(\mathbf{k})|}\right\rangle_{\varepsilon\geq |\Delta(\mathbf{k})|} \nonumber\\
&\approx&\!\! i\frac{\pi \varepsilon}{2\varepsilon^{\star}},
\end{eqnarray}
where $\varepsilon^{\star} = \pi \left[\sum_{j}|u_{\mathbf{k}_j}(\mathbf{r}_0)|^2/ |\Delta^{\prime}_j|\right]^{-1}$, the sum runs over all nodes $\mathbf{k}_j$ of the gap, and $\Delta^\prime =\partial\Delta(\mathbf{k})/\partial\tau$ is the derivative of the gap over a dimensionless vector tangential to the Fermi line.

A similar analysis for the integral in $a_{p,h}$ shows that
\begin{equation}
\label{app-SB-2-Bg}
\left\langle |u_{\mathbf{k}}(\mathbf{r}_0)|^2 \frac{\Delta^{*}(\mathbf{k})}{\sqrt{|\Delta(\mathbf{k})|^2 -\varepsilon^2}}\right\rangle_{\varepsilon\geq |\Delta(\mathbf{k})|} \approx {\cal{O}}\left(\frac{\varepsilon}{\varepsilon^{\star}}\right)^2.
\end{equation}
Unlike Eq.~(\ref{app-SB-2-Ag}), the above integral is nonzero only if $\Delta(\mathbf{k})$ violates the lattice symmetry or if $\mathbf{r}_0$ is not a high-symmetry point.

In the case $\varepsilon\leq |\Delta(\mathbf{k})|$, $a_{ph}/\varepsilon$ is logarithmically divergent at $\varepsilon\to 0$ with the leading asymptote scaling as
\begin{equation}
\label{app-SB-2-Al}
\left\langle \frac{|u_{\mathbf{k}}(\mathbf{r}_0)|^2 \,\varepsilon}{\sqrt{|\Delta(\mathbf{k})|^2-\varepsilon^2}} \right\rangle_{\varepsilon\leq |\Delta(\mathbf{k})|} \approx \frac{\varepsilon}{\varepsilon^{\star}} \ln{\left(\frac{\Delta_{\rm max}}{\varepsilon}\right)}.
\end{equation}
Similar to Eq.~(\ref{app-SB-2-Bg}), the integral over the Fermi line in $a_{p,h}$ at $\varepsilon\leq |\Delta(\mathbf{k})|$ vanishes if the gap $\Delta(\mathbf{k})$ or the tip position $\mathbf{r}_0$ does not violate the respective symmetries. We estimate the corresponding integral as
\begin{eqnarray}
\label{app-SB-2-Bl}
&&\left\langle |u_{\mathbf{k}}(\mathbf{r}_0)|^2 \frac{\Delta^{*}(\mathbf{k})}{\sqrt{|\Delta(\mathbf{k})|^2 -\varepsilon^2}}\right\rangle_{\varepsilon\leq |\Delta(\mathbf{k})|} \nonumber\\
&&\approx \left\langle |u_{\mathbf{k}}(\mathbf{r}_0)|^2 \frac{\Delta^{*}(\mathbf{k})}{|\Delta(\mathbf{k})|} \right\rangle_{0}
+{\cal{O}}\left(\left(\frac{\varepsilon}{\varepsilon^{\star}}\right)^2\ln{\left(\frac{\Delta_{\rm max}}{\varepsilon}\right)}\right).\nonumber\\
\end{eqnarray}

The leading-order expansions given in Eqs.~(\ref{app-SB-2-Ag}), (\ref{app-SB-2-Al}), and (\ref{app-SB-2-Bl}) are used in Sec.~\ref{sec:Resonances}.

\section{Bound states for a \texorpdfstring{$d_{x^2-y^2}+id_{xy}$}{d+id} gap}
\label{sec:App-d+id}

\begin{figure}[!ht]
\centering
\includegraphics[width=0.45\textwidth]{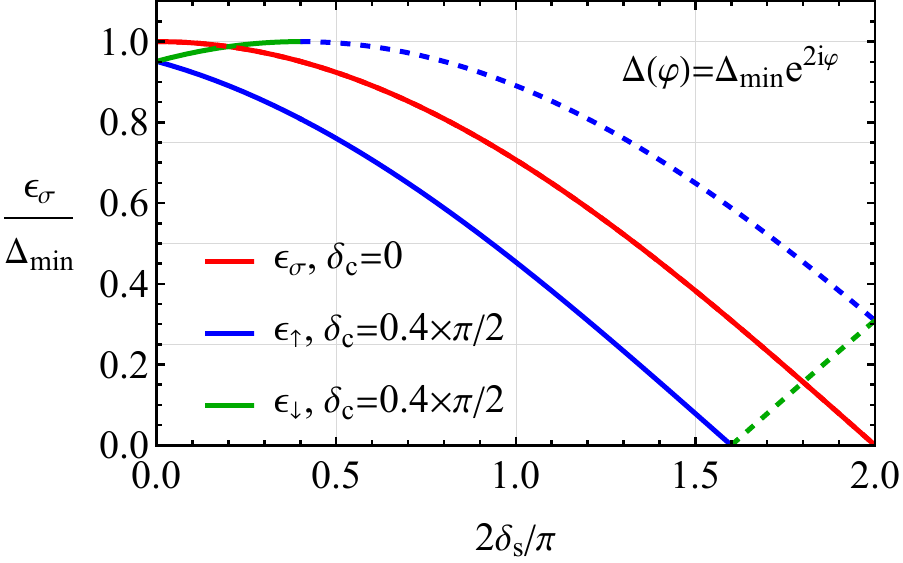}
\caption{
Normalized energy of the subgap states $\epsilon_{\sigma}/\Delta_{\rm min}$ as a function of $\delta_{s}$ for $\delta_{c}=0$ (red line) and $\delta_{c}=0.4\,\pi/2$ (blue and green lines). The presence of spin-dependent scattering splits the spin-degenerate level (red line) into two states $\epsilon_{\uparrow}$ (blue line) and  $\epsilon_{\downarrow}$ (green line) at $\delta_s \leq \delta_c$. The bound states have the same spin polarization for $\delta_s \geq \delta_c$. Solid and dashed lines show two different branches of bound states with the same spin polarization, see $\epsilon_{\uparrow}^{p}$ and $\epsilon_{\uparrow}^{h}$ in Eq.~(\ref{Resonances-sub-TRS-d+id-sol}). We assume a circular Fermi surface (parametrized by the angle $\varphi$), consider tunneling into a high symmetry point, and use the $d_{x^2-y^2}+id_{xy}$ gap $\Delta(\varphi) = \Delta_{\rm min} e^{2i\varphi}$.
}
\label{fig:App-d+id-TRS}
\end{figure}

In this section, we discuss subgap bound states for the $d_{x^2-y^2}+id_{xy}$ gap $\Delta(\varphi) = \Delta_{\rm min} e^{2i\varphi}$ used in Sec.~\ref{sec:Resonances-sub-TRS} in more detail. This gap preserves the lattice point symmetry but, obviously, breaks TRS. The dispersion relation of the subgap states with positive energies can be obtained exactly,
\begin{equation}
\label{Resonances-sub-TRS-d+id-sol}
\epsilon_{\sigma}^{p,h} = \Delta_{\rm min} \cos{\left(\frac{\sigma \delta_s \pm \delta_c}{2}\right)}, \quad \cot{\left(\sigma \delta_s \pm \delta_c\right)} \geq0;
\end{equation}
see Eqs.~(\ref{Iplusminus-symm}) and (\ref{pointsymm-no-TRS}). Here, $\delta_s=\delta_{\uparrow}-\delta_{\downarrow}$ and $\delta_c=\delta_{\uparrow}+\delta_{\downarrow}$. The particle-hole symmetry dictates $\epsilon_{\sigma}^{p,h} = -\epsilon_{-\sigma}^{h,p}$. Perhaps, the most noticeable feature of Eq.~(\ref{Resonances-sub-TRS-d+id-sol}), is the presence of two spin-polarized subgap states due to the spin-dependent scattering if $\delta_s \leq\delta_c$ and two states with the same spin polarization at $\delta_s \geq\delta_c$. The bound states are spin polarized and doubly degenerate at $\delta_c=0$. We illustrate the dispersion relation of the bound states in Fig.~\ref{fig:App-d+id-TRS}, where a pair of spin-polarized states at $\delta_s \leq \delta_c$ evolves into a pair of states with the same spin polarization at $\delta_s \geq \delta_c$; cf. Fig.~\ref{fig:Resonances-sub-TRS}(a). The observability of these states in STS is discussed after Eq.~(\ref{pointsymm-no-TRS}).

\section{Role of scattering potentials for above-gap energies}
\label{sec:App-Resonances-above}

For the sake of completeness, let us address the role of impurity scattering for the above-gap energies $\varepsilon>\Delta_{\rm max} =\mbox{max}{\left\{|\Delta(\mathbf{k})|\right\}}$. While there are no bound states in this case, it is instructive to show how the impurity scattering potential affects the differential conductance.

By using Eqs.~(\ref{Resonances-gen-G-def})--(\ref{Resonances-gen-pha-s-Np}) and Eq.~(\ref{Iplusminus}), and assuming the tunneling limit $|t|^2\ll1$ in which the Andreev scattering amplitudes can be neglected, we obtain the following conductance for above-gap energies:
\begin{eqnarray}
\label{Resonances-above-G}
G_{\sigma}(V) &=& \frac{G_n}{64|\mathcal{D}_{\sigma}|^2} \frac{I_{-}(|eV|)-I_{+}(|eV|)}{i} \nonumber\\
&\times&\Bigg(\left[1-\RE{I_{-}(|eV|)I_{+}^{*}(|eV|)}\right]\cos{(\delta_c-\sigma \delta_s)}\nonumber\\
&+& \RE{I_{-}(|eV|)I_{+}^{*}(|eV|)}+1  \Bigg),
\end{eqnarray}
where the analytical continuation to $\varepsilon>|\Delta(\mathbf{k})|$ in $I_{\pm}(|eV|)$ given in Eq.~(\ref{Iplusminus}) was defined as $\sqrt{-z}=-i\sqrt{|z|}$ at $z>0$. Notice that, for momentum-dependent gaps, the spin degeneracy of the above-gap conductance is lifted. In this case, the last term in the curly brackets may remain nonvanishing if $\RE{I_{-}(|eV|)I_{+}^{*}(|eV|)}\neq 1$.

For the $s$-wave gap $\Delta(\mathbf{k}) = \Delta$, the conductance (\ref{Resonances-above-G}) is simplified as
\begin{equation}
\label{Resonances-above-G-swave}
G_{\sigma}(V)  = \frac{G_n |eV|\sqrt{(eV)^2-\Delta^2}}{(eV)^2 - \Delta^2 \cos^2{(\delta_s/2)}}.
\end{equation}
This conductance is spin-degenerate and is unaffected by potential scattering. At $\delta_s=0$, the conventional expression for impurity-free superconductors with a square-root divergency $\propto 1/\sqrt{(eV)^2-\Delta^2}$ at $|eV|\to\Delta$ is reproduced. On the other hand, for spin-dependent scattering $\cos{(\delta_s/2)}\neq1$, the conductance (\ref{Resonances-above-G-swave}) vanishes as $\propto \sqrt{(eV)^2-\Delta^2}$ at $|eV|\to \Delta$.

\bibliography{library-short}

\end{document}